\documentclass[aps,prx,twocolumn,amsmath,amssymb,superscriptaddress,longbibliography]{revtex4-1}

\usepackage{graphicx} 
\graphicspath{{figures/}}

\usepackage{dcolumn} 
\usepackage{color}
\usepackage{amsmath} 
\usepackage{braket}
\usepackage[breaklinks,colorlinks,bookmarks=false,citecolor=blue,linkcolor=red,urlcolor=blue]{hyperref}
\usepackage{cleveref} 
\crefname{equation}{Eq.}{Eqs.}
\Crefname{equation}{Eq.}{Eqs.}  
\crefname{figure}{Fig.}{Figs.}
\Crefname{figure}{Fig.}{Figs.}  
\crefname{section}{section}{sections}
\Crefname{section}{Section}{Sections}

\DeclareMathOperator{\tr}{Tr}
\newcommand\norm[1]{\left\lVert#1\right\rVert}

\begin{document}

\title{Thermodynamic properties of the Shastry-Sutherland model throughout 
the dimer-product phase}

\author{Alexander Wietek}
\email{awietek@flatironinstitute.org}
\affiliation{Center for Computational Quantum Physics, Flatiron Institute, 
New York, NY 10010, USA}
\affiliation{Institut f\"ur Theoretische Physik, Universit\"at Innsbruck, 
A-6020 Innsbruck, Austria}

\author{Philippe Corboz}
\affiliation{Institute for Theoretical Physics and Delta Institute for 
Theoretical Physics, University of Amsterdam, Science Park 904, 1098 XH 
Amsterdam, The Netherlands}

\author{Stefan Wessel}
\affiliation{Institut f\"ur Theoretische Festk\"orperphysik, JARA-FIT and 
JARA-HPC, RWTH Aachen University, 52056 Aachen, Germany}

\author{B.~Normand}
\affiliation{Neutrons and Muons Research Division, Paul Scherrer Institute, 
CH-5232 Villigen-PSI, Switzerland}

\author{Fr\'ed\'eric Mila}
\affiliation{Institute of Physics, Ecole Polytechnique F\'ed\'erale
Lausanne (EPFL), CH-1015 Lausanne, Switzerland}

\author{Andreas Honecker}
\affiliation{Laboratoire de Physique Th\'eorique et Mod\'elisation, CNRS 
UMR 8089, Universit\'e de Cergy-Pontoise, 95302 Cergy-Pontoise Cedex, France}

\date{\today}

\begin{abstract}
The thermodynamic properties of the Shastry-Sutherland model have posed one 
of the longest-lasting conundrums in frustrated quantum magnetism. Over a 
wide range on both sides of the quantum phase transition (QPT) from the 
dimer-product to the plaquette-based ground state, neither analytical nor 
any available numerical methods have come close to reproducing the physics 
of the excited states and thermal response. We solve this problem in the 
dimer-product phase by introducing two qualitative advances in computational 
physics. One is the use of thermal pure quantum (TPQ) states to augment 
dramatically the size of clusters amenable to exact diagonalization. The 
second is the use of tensor-network methods, in the form of infinite projected 
entangled pair states (iPEPS), for the calculation of finite-temperature 
quantities. We demonstrate convergence as a function of system size in TPQ 
calculations and of bond dimension in our iPEPS results, with complete mutual 
agreement even extremely close to the QPT. Our methods reveal a remarkably 
sharp and low-lying feature in the magnetic specific heat, whose origin 
appears to lie in a proliferation of excitations composed of two-triplon 
bound states. The surprisingly low energy scale and apparently extended 
spatial nature of these states explain the failure of less refined numerical 
approaches to capture their physics. Both of our methods will have broad and 
immediate application in addressing the thermodynamic response of a wide 
range of highly frustrated magnetic models and materials. 
\end{abstract}

\maketitle

\section{Introduction}
\label{sintro}

Frustrated quantum spin systems, particularly those forbidding
magnetically ordered ground states, provide one of the most important
avenues in condensed matter physics for the realization and
investigation of phenomena ranging from fractionalization to
many-particle bound states, from massive degeneracy to topological
order, and from quantum entanglement to quantum criticality
\cite{HFMbook}. Significant progress has been made in describing the
ground states of many such quantum magnets in two and three dimensions
(2D and 3D), including different types of valence-bond crystal and
both gapped and gapless quantum spin liquids \cite{Savary_2016}. However, 
a full understanding of the excitation spectrum and thermodynamic
response of frustrated quantum spin models, which is the key to
experimental interpretation, has remained elusive in all but a small
number of special cases. This is because no unbiased analytical or
numerical methods exist by which to study these properties for general
systems at low but finite temperatures and on lattices whose sizes
approximate the thermodynamic limit.

\begin{figure}[t]
  \includegraphics[width=\columnwidth]{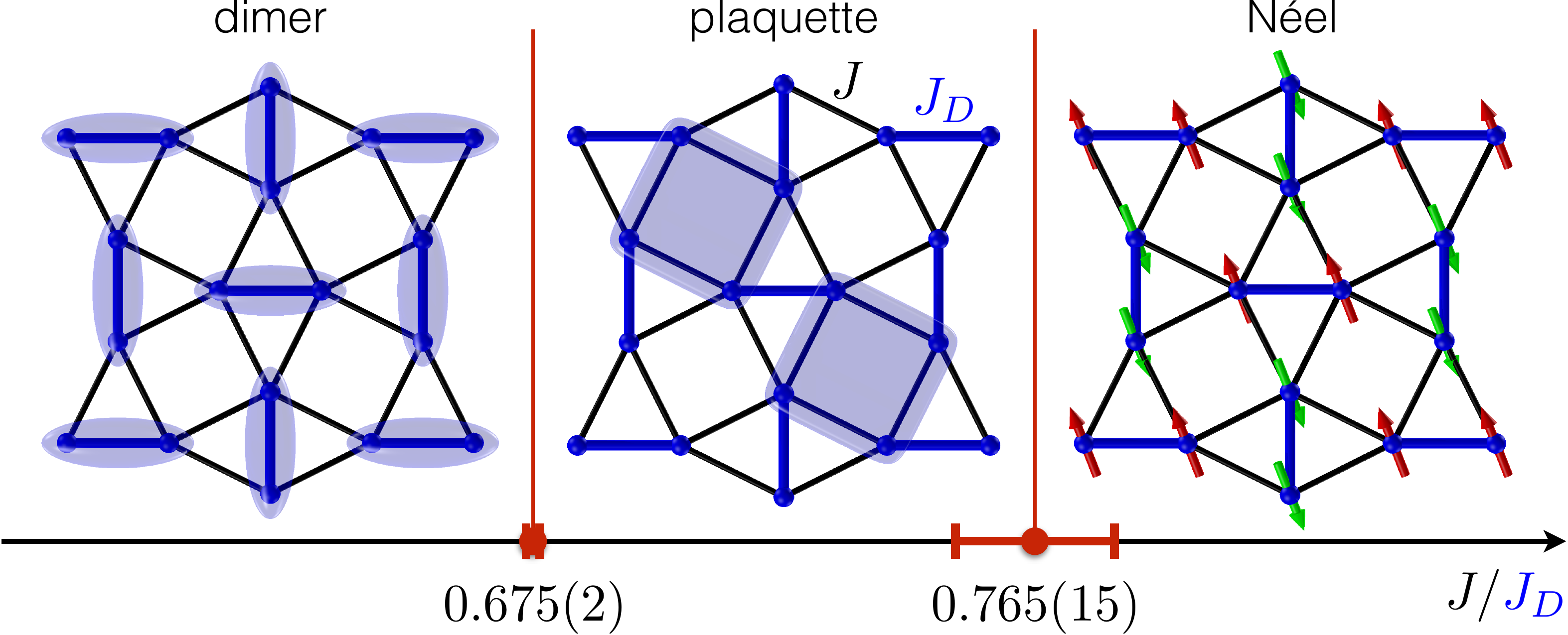}
  \caption{Geometry and coupling constants, $J_D$ and $J$ (center panel), of 
  the Shastry-Sutherland model [Eq.~(\ref{ess})]. The three panels show 
  schematic representations of the singlet dimer-product state, which is 
  exact, the plaquette state, and the N\'eel phase. Shaded ellipses and 
  squares denote respectively dimer- and plaquette-singlet states. Phase 
  boundaries are taken from Ref.~\cite{Corboz13_shastry}.} 
  \label{fig:sspd}
\end{figure}

A case in point is the Shastry-Sutherland model \cite{ShaSu81}, which
was constructed explicitly to have as its ground state a product state
of dimer singlets. Also referred to as the orthogonal dimer model
\cite{MiUeda03}, it is defined by the Hamiltonian
\begin{equation}
  H = J_D \sum_{\langle i,j\rangle} \vec S_i \cdot \vec S_j + J \sum_{\langle \langle 
    i,j\rangle\rangle} \vec S_i \cdot \vec S_j,
  \label{ess}
\end{equation}
which is represented schematically in Fig.~\ref{fig:sspd}. $J_D$ is an 
intra-dimer coupling (denoted by $\langle ij \rangle$) and $J$ a mutually 
frustrating inter-dimer coupling (denoted $\langle \langle ij \rangle 
\rangle$). The ground state is clearly a dimer-product phase at small 
$J/J_D$ and a square-lattice antiferromagnet (AF) at large coupling ratios. 
As intimated above, the ground-state phase diagram is in fact quite well 
known, with the most quantitatively accurate results provided by the ansatz 
of infinite projected entangled-pair states (iPEPS, which we will extend 
here to finite temperatures) \cite{Corboz13_shastry}. As shown in 
Fig.~\ref{fig:sspd}, the singlet dimer-product state is found up to
$J/J_D = 0.675(2)$ and the AF state beyond $J/J_D = 0.765(15)$, with a
gapped, plaquette-based state occurring in the intermediate regime. The 
lower quantum phase transition (QPT) is of first order; the upper one 
has also been found numerically to be first-order, but only weakly so 
\cite{Corboz13_shastry}, and this has also been questioned \cite{Lee2019}.
Any finite temperature will destroy the magnetic order of the AF phase in 
this 2D system. By contrast, in the plaquette phase one expects an Ising 
transition to occur at finite temperatures as the plaquette order, which 
breaks the lattice symmetry, is lost. In the dimer-product state, no such 
physics is expected.

Somewhat remarkably, the Shastry-Sutherland model has a faithful realization 
in the material SrCu$_2$(BO$_3$)$_2$, which has exactly the right geometry and 
only weak non-Heisenberg (primarily Dzyaloshinskii-Moriya) interactions. 
Discovered in 1991 \cite{Smith1991} and first studied for its magnetic 
properties 20 years ago this year \cite{Kageyama1999}, SrCu$_2$(BO$_3$)$_2$ 
lies in the dimer-product phase \cite{MiUeda99} and shows a very flat band of 
elementary triplet (``triplon'') excitations \cite{Kageyama2000b,Gaulin2004}. 
Excellent magnetic susceptibility \cite{Kageyama1999,Kageyama1999b} and 
specific-heat \cite{Kageyama2000} measurements made at this time were used 
to estimate the coupling ratio as $J/J_D \simeq 0.635$, thereby placing 
SrCu$_2$(BO$_3$)$_2$ rather close to the dimer-plaquette QPT. Detailed 
experiments performed in the intervening two decades have revealed a 
spectacular series of magnetization plateaus \cite{Kageyama1999,Onizuka00,
Kodama02,Sebastian08,Takigawa2011,Takigawa12,Jaime12,PhysRevLett.111.137204,
ncomms16}, as well as a curious redistribution of spectral weight at 
temperatures very low on the scale of the triplon gap \cite{Gaulin2004,
Zayed2014}. Of most interest to our current study is the result 
\cite{Zayed2017} that an applied pressure makes it possible to increase 
the ratio $J/J_D$ to the extent that, at approximately 1.9 GPa, the material 
is pushed through the QPT into a plaquette phase. Data for the specific heat 
under pressure have only appeared \cite{Guo2019} concurrently with the present 
study, and indicate that the low-temperature peak moves to a lower temperature 
at 1.1 GPa before evidence of an ordered phase appears at 1.8 GPa. 

In contrast to the ground state, the excited states and thermodynamics
of the Shastry-Sutherland model are not at all well known around the QPT, 
despite the attention focused on this regime due to SrCu$_2$(BO$_3$)$_2$. 
Full exact diagonalization (ED) has been limited to systems of up to 20 
sites and, as we will demonstrate in detail here, suffer from significant 
finite-size limitations at $J/J_D > 0.6$. Quantum Monte Carlo (QMC) methods 
suffer from extreme sign problems, and despite a recent study by some of us
\cite{Wessel2018} that expands the boundaries of what is possible by QMC,
the low-temperature behavior in the regime $J/J_D > 0.6$ remains entirely 
inaccessible. iPEPS methods work in the thermodynamic limit and are immune to 
frustration problems, but to date have been limited to ground-state properties. 
As a consequence, two decades after their measurement, the best interpretation 
of the thermodynamic properties of SrCu$_2$(BO$_3$)$_2$ \cite{Kageyama1999b,
Kageyama2000}, and the origin of the estimate $J/J_D = 0.635$, remain based 
on ED of 16- and 20-site clusters \cite{MiUeda00,MiUeda03}.

\begin{figure}[t]
  \hbox{\hbox{\includegraphics[width=\columnwidth]{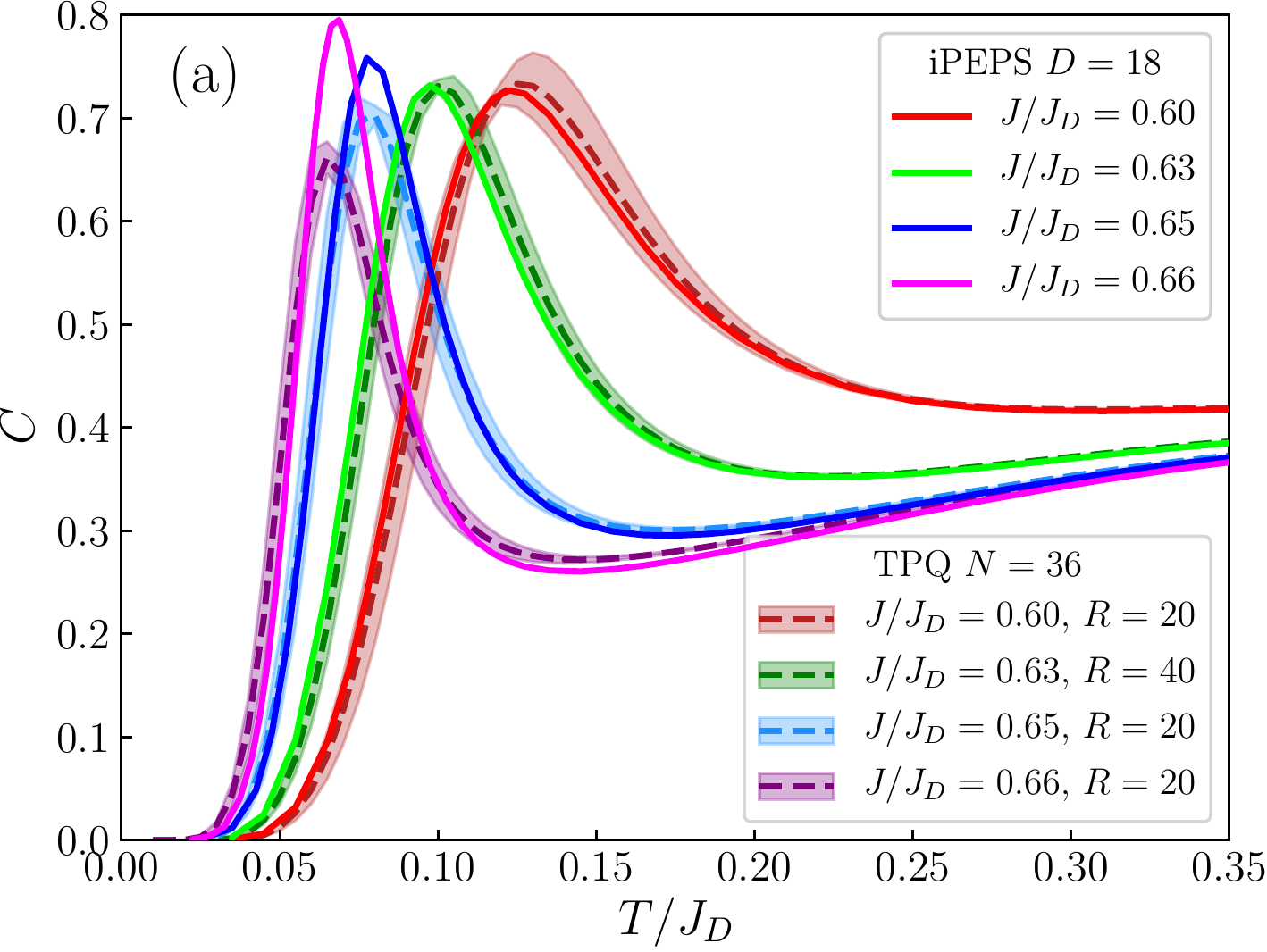}}}
  \hbox{\hbox{\includegraphics[width=\columnwidth]{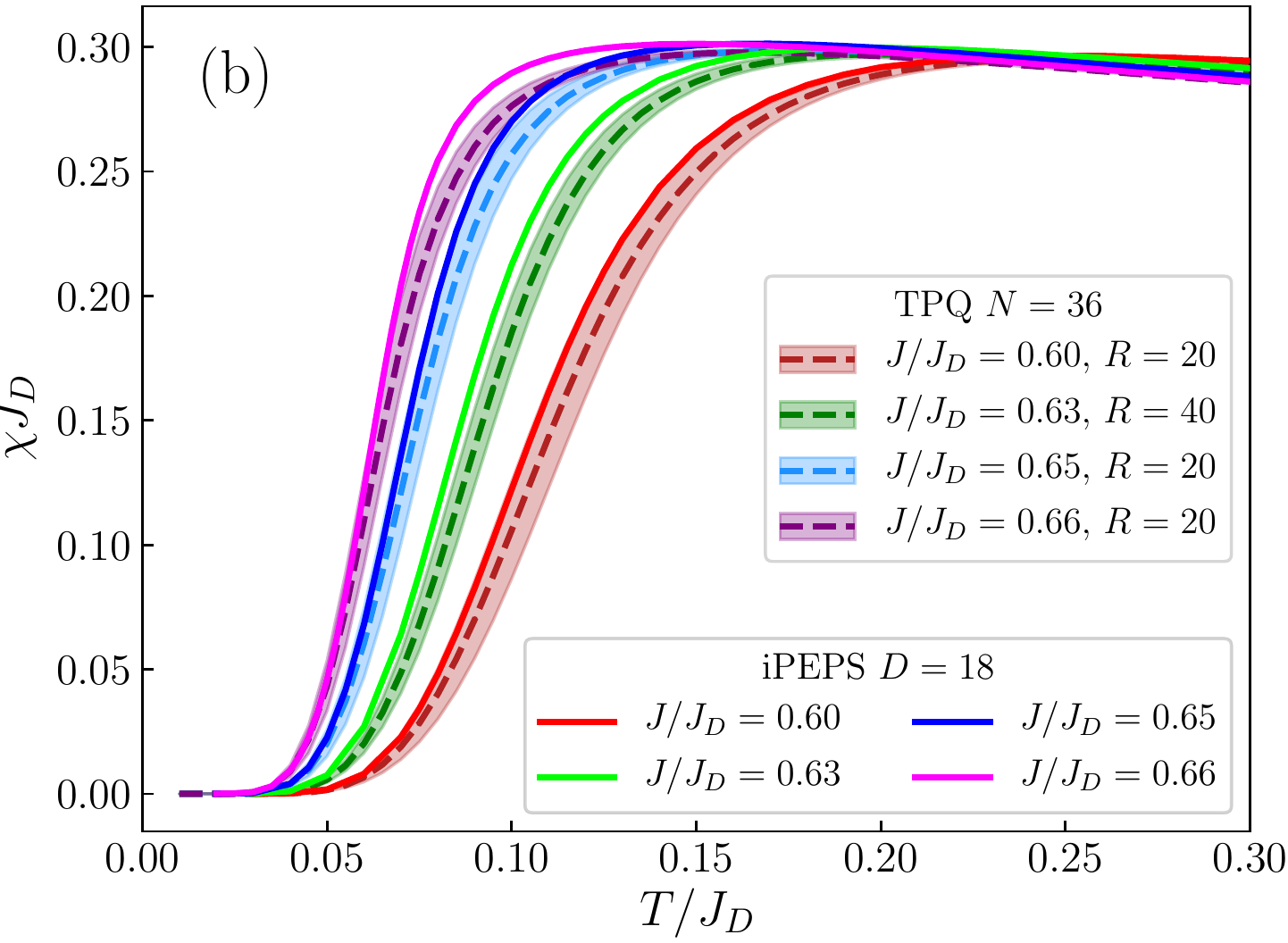}}}
  \caption{(a) Magnetic specific heat, $C(T)$, and (b) magnetic susceptibility,
    $\chi(T)$, of the Shastry-Sutherland model in the regime of coupling ratios 
    $0.60 \le J/J_D \le 0.66$. The number of sites in the TPQ calculations is 
    $N = 36$ and the shaded regions show the standard error of the TPQ method 
    for the respective coupling ratios (Sec.~\ref{tresults}). The iPEPS bond 
    dimension is $D = 18$.}
  \label{fig:sspdmultiple}
\end{figure}

Here we introduce two qualitative technical advances in numerical methods 
for computing thermodynamic properties and benchmark them by application to 
the Shastry-Sutherland model in the dimer-product phase up to the QPT. For a 
quantum many-body system, a thermal equilibrium state can be represented by 
a typical pure state, which is known as a thermal pure quantum (TPQ) state 
\cite{Sugiura2012,Sugiura2013}. This result can be understood as a consequence 
of a more general phenomenon, known as quantum typicality~\cite{Hams2000, 
Goldstein2006}. All statistical-mechanical quantities can be 
obtained from averaging over a few TPQ states. For practical purposes, 
obtaining TPQ states by ED allows access to thermodynamic quantities without 
having to perform a full diagonalization, and hence admits a qualitative 
increase in accessible system sizes \cite{Yamaji2014,Yamaji2016}. We will 
exploit TPQ methods to reach system sizes of $N = 40$ in the Shastry-Sutherland 
problem.

iPEPS are a tensor-network ansatz to represent a quantum state on an
infinite lattice~\cite{verstraete2004,jordan2008,nishio2004}, here in 2D, 
whose accuracy is controlled by the bond dimension, $D$, of the tensor. In 
fact this ansatz provides not only a compact representation for the ground 
states of gapped local Hamiltonians, but also for representing a purification 
of the thermal density operator, and hence the thermal states of local 
Hamiltonians. Here we exploit newly developed algorithms~\cite{czarnik19} 
allowing the efficient calculation of imaginary-time evolution processes, which 
in general require additional truncation to retain the value of $D$ at each 
time step, to generate thermal states of the Shastry-Sutherland Hamiltonian
and hence to compute its thermodynamic properties. 

Our TPQ and finite-$T$ iPEPS methods enable us to capture the physics of 
the Shastry-Sutherland model, and thus of SrCu$_2$(BO$_3$)$_2$, in a way 
not hitherto possible near the QPT. To illustrate this clearly, in 
Fig.~\ref{fig:sspdmultiple} we show sample results from both methods for 
the magnetic specific heats and susceptibilities of Shastry-Sutherland 
models with coupling ratios $J/J_D = 0.60$, 0.63, 0.65, and 0.66. Over 
this small range of couplings near the QPT, the specific heat develops a 
pronounced two-peak form, with a broad higher peak lying just beyond the 
range of Fig.~\ref{fig:sspdmultiple}(a). The narrow lower peak, presumably 
also characteristic of finite-energy excitations in the magnetically 
disordered dimer-product phase, becomes surprisingly sharp and moves 
to a remarkably low energy as the QPT is approached. The corresponding 
susceptibilities rise increasingly abruptly before an increasingly ``square'' 
turnover at their maximum. We draw attention to the fact that the results 
from both our methods are in excellent quantitative agreement, apart from the 
specific-heat peak heights extremely close to the QPT. These characteristic 
shapes, and their systematic evolution, have not been obtained before and 
shed new light both on the thermodynamic quantities themselves and on the 
underlying excitation spectrum responsible for their form.

The manuscript is organized as follows. In Sec.~\ref{ttpq} we introduce 
TPQ states and their application to the calculation of thermodynamic 
quantities. In Sec.~\ref{tipeps} we provide a brief introduction to iPEPS 
and discuss the generation of thermal states by their imaginary-time 
evolution. Section \ref{tresults} presents our complete results from 
both methods for the magnetic specific heat and susceptibility of the 
Shastry-Sutherland model over the coupling range $0.60 \le J/J_D \le 0.66$, 
which are summarized in Fig.~\ref{fig:sspdmultiple} and which it has not 
been possible to obtain with any reliability in previous studies. In 
Sec.~\ref{bla} we discuss the physics of the sharp features we observe in 
the specific heat near the QPT, the limits of our numerical methods in the 
context of the Shastry-Sutherland model, and the application of our results 
to the material SrCu$_2$(BO$_3$)$_2$. Section \ref{wrap} contains a brief
summary and perspective concerning the impact of our results in frustrated 
quantum magnetism.

\section{Thermodynamics from thermal pure quantum states}
\label{ttpq}

To evaluate the thermal average of a quantum mechanical observable,
$A$, in the canonical ensemble, one computes
\begin{equation}
  \label{eq:ensembleaverage}
  \langle A \rangle = \tr ({\rm e}^{-\beta H} A) / \mathcal{Z},
\end{equation}
where $H$ denotes the Hamiltonian, $\beta = 1/ (k_{\textrm{B}}T)$ the
inverse temperature, $k_{\textrm{B}}$ the Boltzmann constant, and
$\mathcal{Z} = \tr ({\rm e}^{-\beta H})$ the canonical partition
function.  Thermal pure quantum (TPQ) states provide an alternative
approach to the evaluation of thermodynamic
properties~\cite{Sugiura2012,Sugiura2013}.  The trace of any operator,
$A$, can be evaluated by taking the average of its expectation values
over a set of random vectors,
\begin{equation}
  \label{eq:generictraceaverage}
  \tr(A) = d \, \overline{\braket{r | A | r}},
\end{equation}
where $\ket{r}$ denotes a normalized vector with random
normal-distributed components and $d$ denotes the dimension of the
Hilbert space. Hence, any thermal average can be estimated by
evaluating
\begin{equation}
  \label{eq:tpqaveraging}
  \langle A \rangle = \frac{\overline{\braket{\beta | A | \beta}}}
  {\overline{\braket{\beta | \beta}}},
\end{equation}
where
\begin{equation}
  \label{eq:tpqstate}
  \ket{\beta} \equiv \textrm{e}^{-\frac{\beta}{2} H}  \ket{r}
\end{equation}
denotes the ``canonical TPQ'' state~\cite{Sugiura2013} and
$\overline{\mbox{\dots\raisebox{0.2cm}{}}}$ denotes the average over
the random vectors $\ket{r}$. Importantly, the TPQ state,
$\ket{\beta}$, is still random. The variance of the estimate in
\cref{eq:tpqaveraging} decreases as $1/\sqrt{d}$, where $d = 2^N$ for
a $S = 1/2$ lattice model on $N$ sites, and hence becomes
exponentially small as a function of the system size under only mild
assumptions about the form of the operator $H$; a precise statement
may be found in Ref.~\cite{Sugiura2013}. For larger system sizes, it
is therefore necessary to generate only relatively few random TPQ
states, $\ket{\beta}$, in order to achieve small statistical
errors. Because the samples in the denominator and the numerator of
\cref{eq:tpqaveraging} are correlated when evaluated from the same set
of random vectors, $\ket{r}$, we perform a jackknife
analysis~\cite{Efron1981} to estimate the error bars.

The key advantage of using TPQ states, when compared with evaluating
statistical averages using the standard formalism of
\cref{eq:ensembleaverage}, is that the state $\ket{\beta}$ can be
approximated numerically to arbitrary precision without resort to a
full diagonalization of the Hamiltonian, $H$.  Variants of the TPQ
method have been employed in several numerical studies of static and
dynamical observables~\cite{Yamaji2014,Yamaji2016,Steinigeweg2015,
Steinigeweg2016,Shimokawa2016,Rousochatzakis2018,Laurell2019}. Whereas
Ref.~\cite{Sugiura2013} proposed a Taylor-expansion technique, here we
improve the speed of convergence by using the Lanczos basis
\cite{Lanczos1950} of the Krylov space of $H$ to evaluate the states
$\ket{\beta}$.

The $n$th orthonormal basis vector in the Lanczos basis,
$\ket{v_{n}}$, is given recursively \cite{Allaire2008} by
\begin{align}
  \label{eq:lanczosrecursionabbrev}
  &\ket{v_{n}} = \ket{\hat{v}_{n}}/{b_{n}}, \text{ with } \\
  &\ket{\hat{v}_{n}} = H \ket{v_{n-1}} -
    a_{n-1}\ket{v_{n-1}} - b_{n-1}\ket{v_{n-2}},
\end{align}
where $\ket{v_1} = \ket{r}$, $b_1 = 0$, and
\begin{equation}
  \label{eq:lanczosalphabeta}
  a_n = \braket{v_n | H | v_n} , \quad 
  b_n = \norm{\hat{v}_{n}}.
\end{equation}
By defining the matrix of Lanczos vectors
\begin{equation}
  \label{eq:lanczosmatrixdef}
  V_n = \left( v_1 | \dots | v_{n} \right ),
\end{equation}
the $n$th Lanczos approximation of $H$ can be written as
\begin{equation}
  \label{eq:lanczosapproximation}
  H \approx V_n T_n V_n^\dagger, 
\end{equation}
where $T_n$ is a tridiagonal matrix given by
\begin{equation}
  \label{eq:tmatrix}
  T_n =
  \begin{pmatrix}
    a_1 & b_2  & 0        &       & \cdots & 0 \\
    b_2  & a_2 & b_3  & 0      &   & \vdots  \\
    0        & b_3  & \ddots   &        &        & \\
    &          &          &        & \ddots & 0 \\
    \vdots   &          &          & \ddots & a_{n-1} & b_{n}\\
    0   &  \cdots &   & 0       & b_{n}       & a_n \\
  \end{pmatrix} \! .
\end{equation}
We use this approximation to evaluate
\begin{align}
  \label{eq:finallanczosapprox}
  \begin{split}
    \braket{\beta | A | \beta} &=
    \braket{r |\textrm{e}^{-\frac{\beta}{2} H} A \textrm{e}^{-\frac{\beta}{2} H}| r} \\
    &\approx \braket{r | V_n \textrm{e}^{-\frac{\beta}{2} T_n}
      V_n^\dagger
      A V_n\textrm{e}^{-\frac{\beta}{2} T_n} V_n^\dagger| r} \\
    &= e_1 \textrm{e}^{-\frac{\beta}{2} T_n} V_n^\dagger A
    V_n\textrm{e}^{-\frac{\beta}{2} T_n} e_1,
  \end{split}
\end{align}
where $e_1 = (1, 0, \ldots, 0)^{\textrm{T}}$. In the case that $A$ is
a power of the Hamiltonian, $A = H^k$, we have
\begin{equation}
  \label{eq:appoxhsimplified}
  V_n^\dagger H^k V_n = T_n^k.
\end{equation}
We stress that the only quantities to be computed are the sequences
$a_n$ and $b_n$. Further, results at all temperatures can be obtained
from a single evaluation of the Lanczos basis. The exponentiated
tridiagonal matrices, $\textrm{e}^{-\frac{\beta}{2} T_n}$, are
computed by the eigendecomposition of the matrices $T_n$ and
exponentiation of the diagonal matrix of eigenvalues.  The
eigendecomposition of $T_n$ is computed using an implicit QR method
\cite{demmel1997applied,golub2012matrix,parlett1998symmetric} by
calling the LAPACK routine
\texttt{dsteqr}~\cite{anderson1999lapack,bai2000templates}.  The
dimension, $n$, of the Krylov space is increased until the desired
precision is achieved. The computations in this manuscript required a
dimension $n < 150$ to achieve a precision superior to the statistical
errors.

In our computations we use a block-diagonal form of the Hamiltonian,
\begin{equation}
  \label{eq:hamiltoniandirectsum}
  H = \bigoplus_\rho H_\rho,
\end{equation}
where the blocks $H_\rho$ correspond to the Hamiltonian in different
lattice-symmetry and total-$S_z$ sectors. Statistical averages are
evaluated in each individual block and then summed,
\begin{equation}
  \label{eq:compose}
  \tr\left({\rm e}^{-\beta H} A\right) = \sum_\rho \tr\left({\rm e}^{-\beta H_\rho} 
    A_\rho\right).
\end{equation}
We employ the techniques proposed in Ref.~\cite{Wietek2018} to perform
the computations in the individual symmetry sectors.

It has recently become possible to compute Lanczos approximations in
particular symmetry sectors for systems of up to $N = 50$ $S = 1/2$
sites for the Heisenberg model on the square lattice and $N = 48$
$S = 1/2$ on the kagome lattice~\cite{Wietek2018}. It is worth noting
here that the latter geometry and model are particularly favorable for
TPQ \cite{Sugiura2013} because the density of states is high at very
low energies \cite{LSS11}. However, deriving thermodynamic quantities
from the TPQ method requires that these Lanczos approximations be
computed multiple times in each symmetry sector for a number, $R$, of
random ``seeds'', i.e.~independent random realizations of the vector
$\ket{r}$ in Eq.~(\ref{eq:tpqstate}), and it is $R$ that determines
the statistical error. Thus restrictions on CPU time currently limit
the practical application of the TPQ approach to an upper limit of
40--42 $S = 1/2$ spins. More specifically, the calculations in this
manuscript were performed with $R = 80$ seeds for all coupling ratios
$J/J_D$ on clusters of $N = 32$ spins and $R = 20$ for the $N = 36$
cluster, except at $J/J_D = 0.63$, where $R = 40$ seeds were
evaluated. For the $N = 40$ cluster we used $R = 5$. In total, the TPQ
calculations in this manuscript required approximately 5.5M CPU hours.

We comment that our implementation of the TPQ method is reminiscent of
the finite-temperature Lanczos method \cite{Jaklic1994}, which has
also recently been pushed to a cluster of $N = 42$ $S = 1/2$ spins for
the kagome lattice \cite{SSR18}. Although the precise relation between
these two different methods remains to be understood, currently we
believe that the TPQ approach has a more systematic theoretical
foundation and as a consequence that the errors within this approach
are better controlled.

For system sizes $N \le 20$, full diagonalization of the Hamiltonian
yields numerically exact results with no statistical errors. An
intermediate method is to compute a large number of low-lying
eigenstates, for example by using the variant of the Lanczos method
outlined in Ref.~\cite{HW09}, and to approximate the low-temperature
thermodynamics using these. Despite the significant advantage of also
providing results that are entirely free of statistical errors, this
method remains restricted by the fact that the required numbers of
eigenvalues are large even at the surprisingly low temperatures where
the dominant physical processes occur in the present case
(Sec.~\ref{tresults}). Thus we present benchmarking Lanczos ED results
for cluster sizes up to $N = 28$, although we comment that $N = 32$
would be accessible were we to invest CPU times comparable to those
used in our TPQ calculations.

\begin{figure}[t]
  \includegraphics[width=\columnwidth]{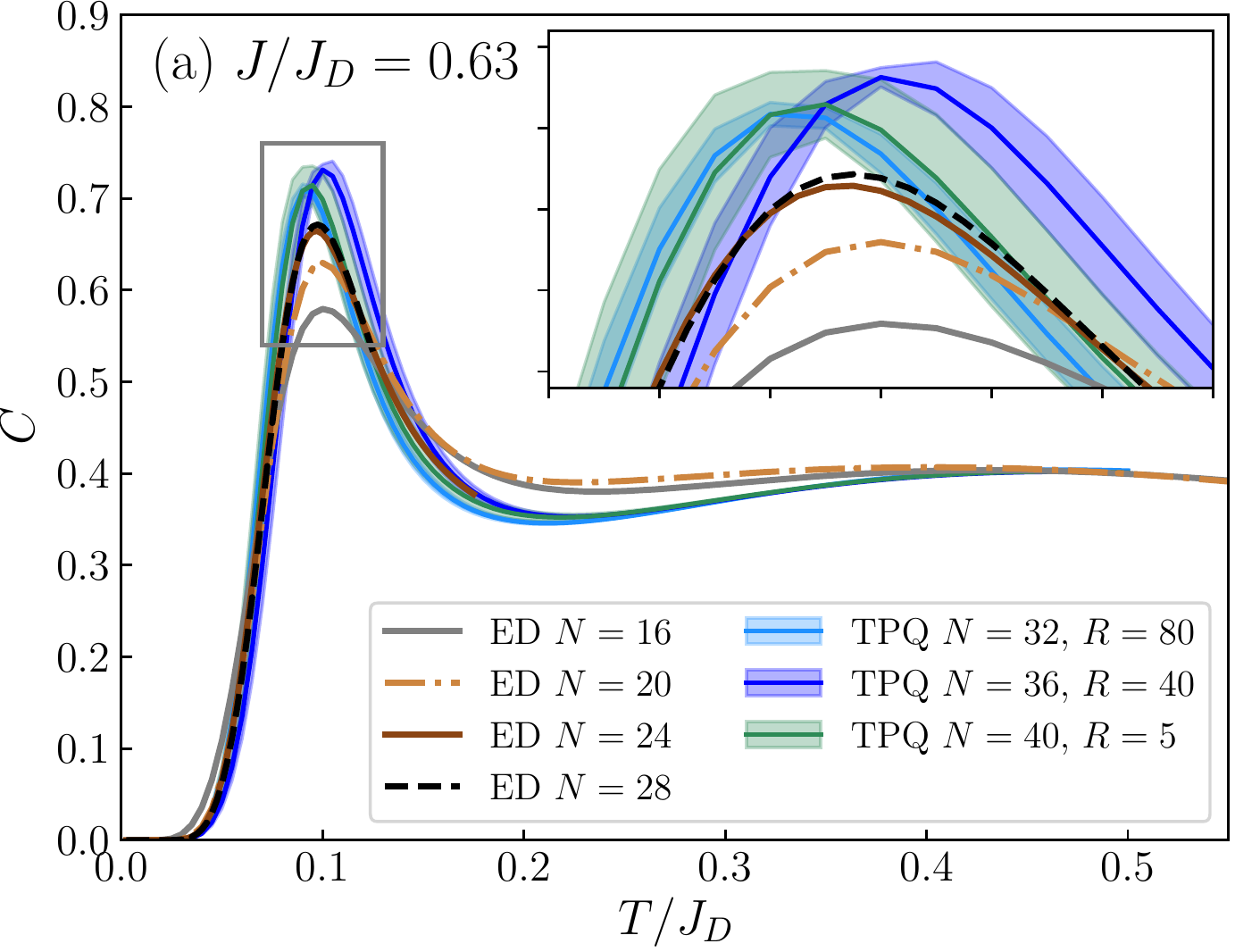}
  \includegraphics[width=\columnwidth]{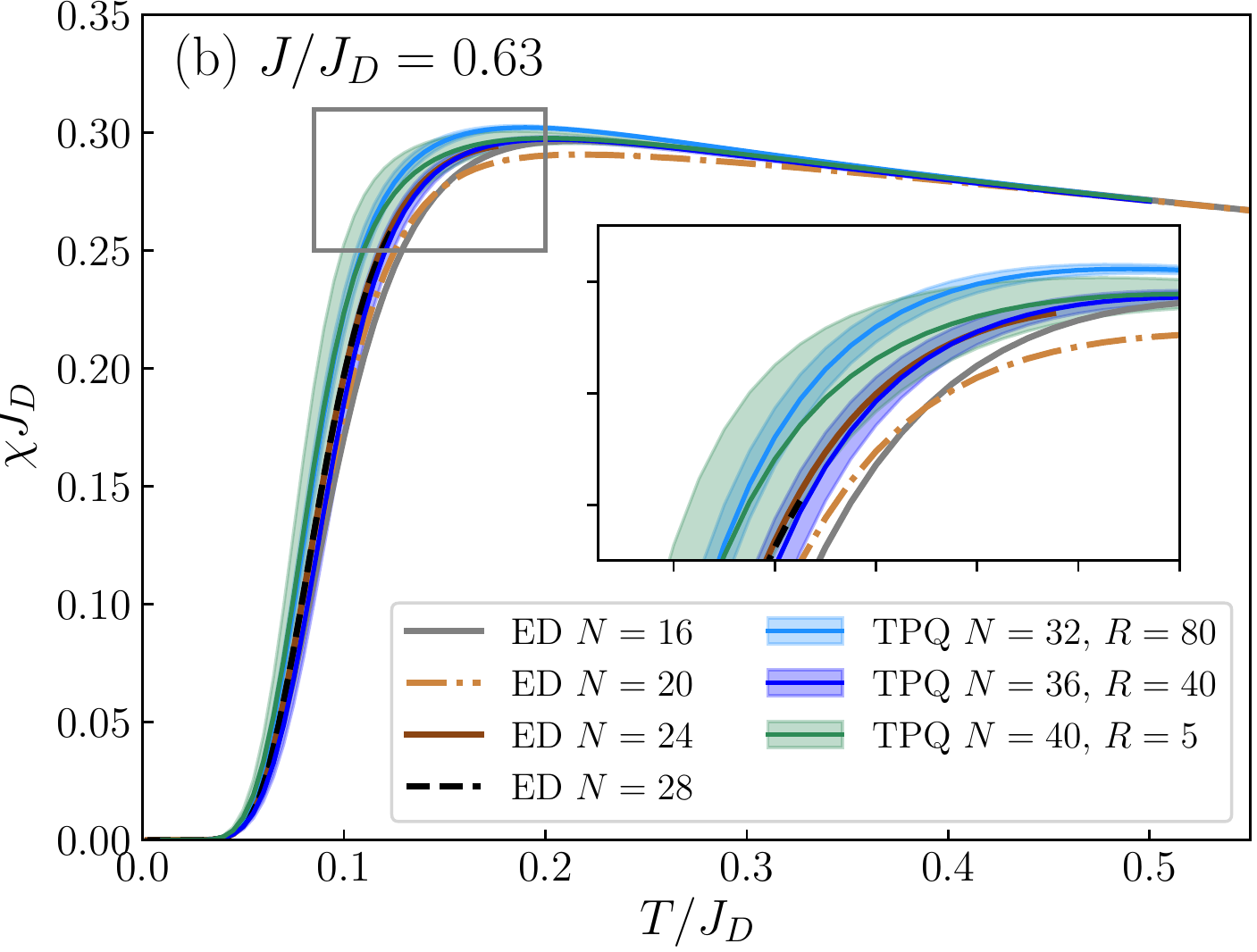}
  \caption{(a) Magnetic specific heat, $C(T)$, and (b) susceptibility,
    $\chi(T)$, per dimer at coupling ratio $J/J_D = 0.63$, comparing
    results obtained from ED and TPQ calculations for system sizes
    from $N = 16$ to 40 sites}
  \label{fig:tpqt}
\end{figure}

As an illustration of the manner in which TPQ system sizes allow
access to the thermodynamic properties of the Shastry-Sutherland
model, in Fig.~\ref{fig:tpqt}(a) we show the specific heat per dimer,
\begin{equation}
  C(T) \equiv  \frac2N \, \frac{\partial E}{\partial T}
  = \frac2N \, \beta^2\left[\braket{H^2} - \braket{H}^2 \right] \!,
  \label{ecdef}
\end{equation}
where $E \equiv \braket{H}$ is the energy, obtained from our TPQ
calculations for coupling ratio $J/J_D = 0.63$ with $N = 32$, 36, and
40. Figure \ref{fig:tpqt}(b) shows the corresponding magnetic
susceptibility per dimer,
\begin{equation}
  \chi(T) \equiv  \frac2N\,\left.\frac{\partial m}{\partial h}\right|_{h=0}
  = \frac2N \, \beta\left[\braket{M^2} - \braket{M}^2\right] ,
  \label{echidef}
\end{equation}
where $h$ is a magnetic field applied along the $z$-axis and
$m \equiv \braket{M}$, with
\begin{equation}
  M = \sum\limits_{i=1}^N S_i^z,
\end{equation}
denotes the total magnetization. In both panels of
Fig.~\ref{fig:tpqt}, we compare these TPQ results with ED calculations
for systems of sizes $N = 16$ and 20, and at low temperatures $N = 24$
and 28. The details and context of these results will be discussed in
full in Secs.~\ref{tresults} and \ref{bla}; for the purposes of this
introduction, we stress that both $C(T)$ and $\chi(T)$ have sharp
physical features, namely the low-temperature peak in the former and
shoulder in the latter, whose shape is manifestly not well reproduced
at small $N$ but becomes very much clearer as $N$ is increased. From
Fig.~\ref{fig:sspdmultiple}, this type of access to larger $N$ values
becomes increasingly important as the QPT is approached.

\section{Thermodynamics from iPEPS}
\label{tipeps}

An infinite projected entangled-pair state (iPEPS), as a tensor-network 
ansatz representing a 2D quantum state in the thermodynamic limit 
\cite{verstraete2004,jordan2008,nishio2004}, can be seen as a natural 
generalization of infinite matrix-product states (MPS) to two, or indeed 
higher, dimensions. While the rank of the tensors used in the representation 
is determined by the physical problem, the amount of information they contain 
is determined by their bond dimension, $D$, which is used to control the 
accuracy of the ansatz. Although iPEPS were introduced originally for 
representing the ground states of local Hamiltonians, more recently several 
iPEPS methods have been developed for the representation of thermal states
\cite{li11,czarnik12,czarnik14,czarnik15,czarnik15b,czarnik16,czarnik16b,
czarnik17,dai17,czarnik18,kshetrimayum18,czarnik19}. Here we focus on the 
approaches discussed in Ref.~\cite{czarnik19}, where an iPEPS is used to 
represent a purification of the 2D thermal density operator. However, we 
note that an alternative route to the calculation of finite-temperature 
properties would be by a direct contraction of the associated (2+1)D tensor 
network using (higher-order) tensor renormalization-group methods~\cite{xie12}.
We comment that MPS-based methods have also been applied recently 
\cite{bruognolo17,chen18,chen19,li19} to compute the thermal response of 
finite 2D systems in cylindrical geometries; these approaches provide 
accurate results for cylinders up to a circumference $W$, which is limited 
by an exponential increase of the required bond dimension with $W$. 
 
\begin{figure}[t]
  \includegraphics[width=\columnwidth]{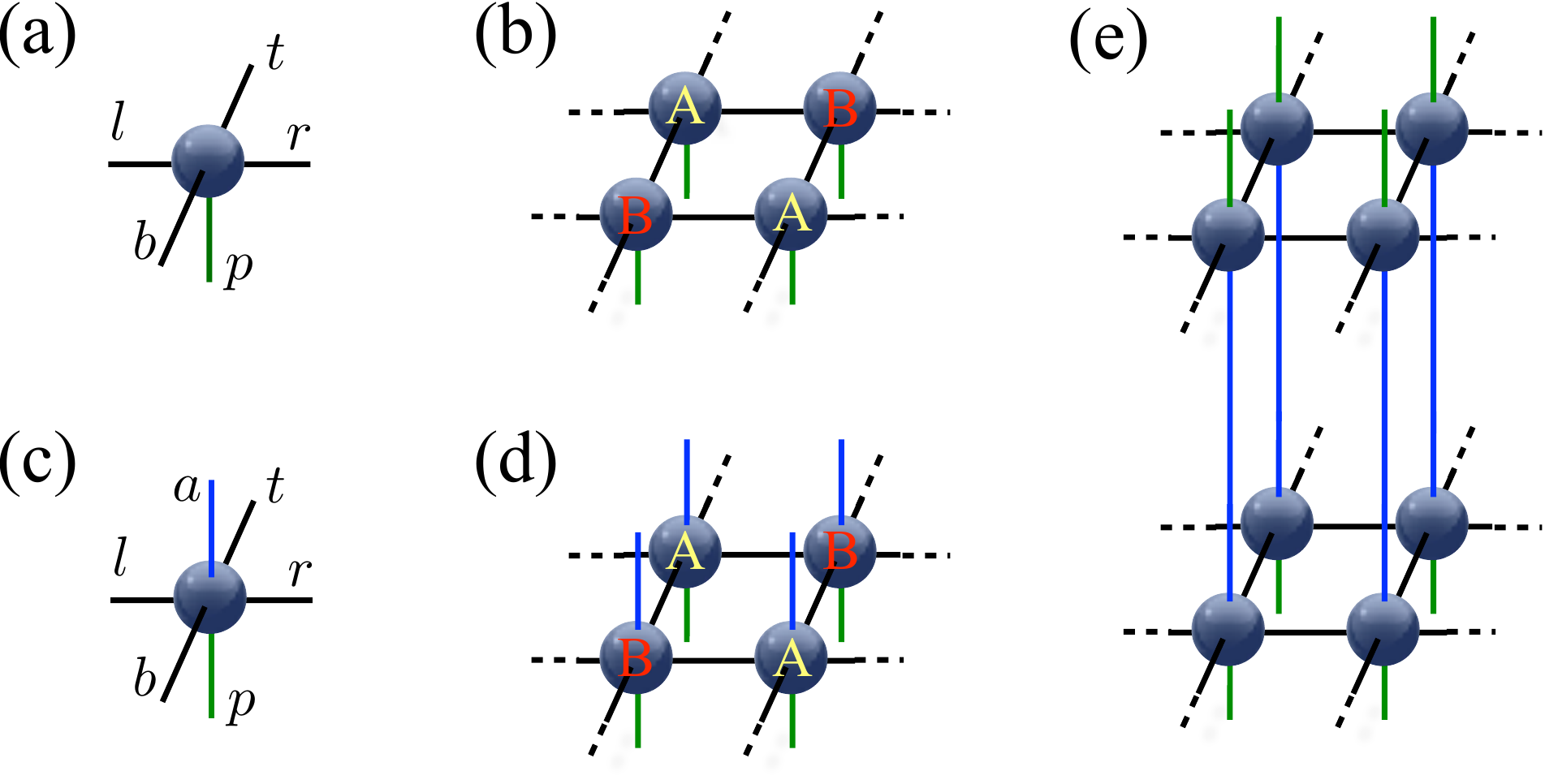}
  \caption{(a)~Graphical representation of an iPEPS tensor, $A_p^{rtlb}$, in 
    which each leg corresponds to an index of the tensor. The physical index, 
    $p$ (green leg), represents the local Hilbert space of a lattice site 
    (here a dimer) and the other four, $r$, $t$, $l$, and $b$ are 
    auxiliary indices, each with bond dimension $D$, which connect neighboring 
    tensors. (b)~A standard iPEPS representing a pure state, shown here for a 
    unit cell containing two different tensors, $A$ and $B$, arranged in a 
    checkerboard pattern on the infinite lattice. A connection between two 
    tensors implies taking the sum over the corresponding index. (c)~iPEPS 
    tensor with an extra index, $a$ (blue leg), representing the local Hilbert 
    space of an ancilla site. (d)~iPEPS ansatz for the purification, 
    $\ket{\Psi(\beta)}$, of the thermal density operator, $\rho(\beta)$. 
    (e)~Representation of $\rho(\beta) = \mathrm{Tr}_{\bf a} \ket{\Psi(\beta)} 
    \bra{\Psi(\beta)}$, where the trace is taken over all ancilla degrees of 
    freedom (shown by the connected blue lines).} 
  \label{fig:iPEPS}
\end{figure}

An iPEPS consists of a unit cell of tensors which is repeated periodically
on an infinite lattice. For a ground state (pure state), each tensor has one 
physical index, $p$, describing the local Hilbert space of the dimer and four 
auxiliary indices, $r$, $t$, $l$, and $b$ [Fig.~\ref{fig:iPEPS}(a)], each with 
bond dimension~$D$, which connect the tensor to its four nearest neighbors on 
a square lattice. For the Shastry-Sutherland model [Fig.~\ref{fig:sspd}] we 
require one tensor per dimer, arranged in a unit cell with two tensors, A and 
B, in a checkerboard pattern, as shown in Fig.~\ref{fig:iPEPS}(b).

We represent a thermal density operator, $\rho(\beta) = e^{-\beta \hat H}$, at 
inverse temperature $\beta$ by its ``purification,'' $\ket{\Psi(\beta)}$, 
which is a pure state in an enlarged Hilbert space where each physical site, 
$j$ (here a dimer with local dimension $d = 4$), is accompanied by an ancilla 
site with the same local dimension, described by a local basis $\ket{p,a}_j$, 
where $p$ and $a$ denote respectively the local physical and ancilla basis 
states. The representation of the thermal state is through its thermal 
density operator, $\rho(\beta) = \mathrm{Tr}_{\bf a} \ket{\Psi(\beta)} \! 
\bra{\Psi(\beta)}$, where the trace is taken over all ancilla degrees of 
freedom, ${\bf a}$. Physically, the ancilla states act as a perfect heat bath, 
and taking the trace gives exact thermodynamic averages. In tensor-network 
notation, each ancilla site is incorporated through the additional index 
$a$ [blue lines in Fig.~\ref{fig:iPEPS}(c)], leading to the diagrammatic 
representation of $\ket{\Psi(\beta)}$ shown in Fig.~\ref{fig:iPEPS}(d) and 
of $\rho(\beta)$ in Fig.~\ref{fig:iPEPS}(e) \footnote{Alternatively, one 
may also regard this iPEPS ansatz as a direct representation of a thermal 
density operator, $\rho(\beta/2)$, at inverse temperature $\beta/2$, where 
each site has both an incoming and an outgoing physical index, and 
Fig.~\ref{fig:iPEPS}(e) corresponds to $\rho(\beta) = \rho^\dagger(\beta/2) 
\rho(\beta/2)$.}.
  
At infinite temperature ($\beta = 0$), $\ket{\Psi(0)}$ can be represented by 
the product state, $\ket{\Psi(0)} = \prod_j \sum_{k=1}^d \ket{k, k}_j$ (with 
bond dimension $D = 1$), which for $\rho(0)$ yields the identity operator. 
The purified state at any given $\beta$ can be obtained as $\ket{\Psi(\beta)}
 = \hat U(\beta/2) \ket{\Psi(0)}$, where the imaginary-time evolution operator 
$\hat U(\beta) = e^{-\beta\hat H}$ acts on the physical degrees of freedom of 
$\ket{\Psi(0)}$. In practice we use a Trotter-Suzuki decomposition of $\hat 
U(\beta)$, meaning it is split into $M$ imaginary-time steps, $\tau = \beta/M$, 
and approximate the operator for a single time step by a product of 
nearest-neighbor time-evolution operators, $\hat U(\tau) = \prod_{\langle i,j 
\rangle} \hat U_{ij}(\tau)$, with $\hat U_{ij}(\tau) = e^{-\tau \hat H_{ij}}$, where 
$\langle i,j \rangle$ denotes all nearest-neighbor site pairs in the unit 
cell~\footnote{Specifically, we employ a second-order Trotter-Suzuki 
decomposition in which the sequence of two-body time-evolution operators 
is reversed at every other time step.}.

Each application of $\hat U_{ij}$ increases the bond dimension between the 
tensors on sites $i$ and $j$ to a larger value, $D' > D$, which for reasons 
of efficiency must be truncated back to $D$ at every step. As in the 
imaginary-time evolution of ground states, there exist several approaches 
for the truncation of a bond index, the two most common being the simple- 
\cite{jiang2008} and full-update methods~\cite{jordan2008,corboz2010}. In 
the simple-update approach, the truncation is performed using a local
approximation of the state, which is computationally cheap but does not 
yield the optimal truncation. In the full-update approach, a bond is 
truncated in a manner optimal under the method by which the entire state 
is taken into account; however, this is considerably more computationally 
expensive and in general it is not possible to reach bond dimensions as 
large as by the simple-update technique. Below we compare results obtained 
by these two approaches. Further details concerning the iPEPS ansatz and 
algorithms for thermal states may be found in Ref.~\cite{czarnik19}.

For contraction of the thermal-state tensor network, which is required to 
compute all physical expectation values, we use a variant~\cite{corboz14_tJ} 
of the corner-transfer-matrix (CTM) renormalization-group 
method~\cite{nishino1996,orus2009-1} in which the accuracy is controlled by 
the boundary bond dimension, $\tilde \chi$. To improve the efficiency of the 
calculations we exploit the global U(1) symmetry \cite{singh2010,bauer2011} 
of the model of \cref{ess}. Further details of the iPEPS methods we employ 
here may be found in Refs.~\cite{corboz2010,Corboz13_shastry,phien15}.

\begin{figure}[t]
  \begin{center}
    \includegraphics[width=1\columnwidth]{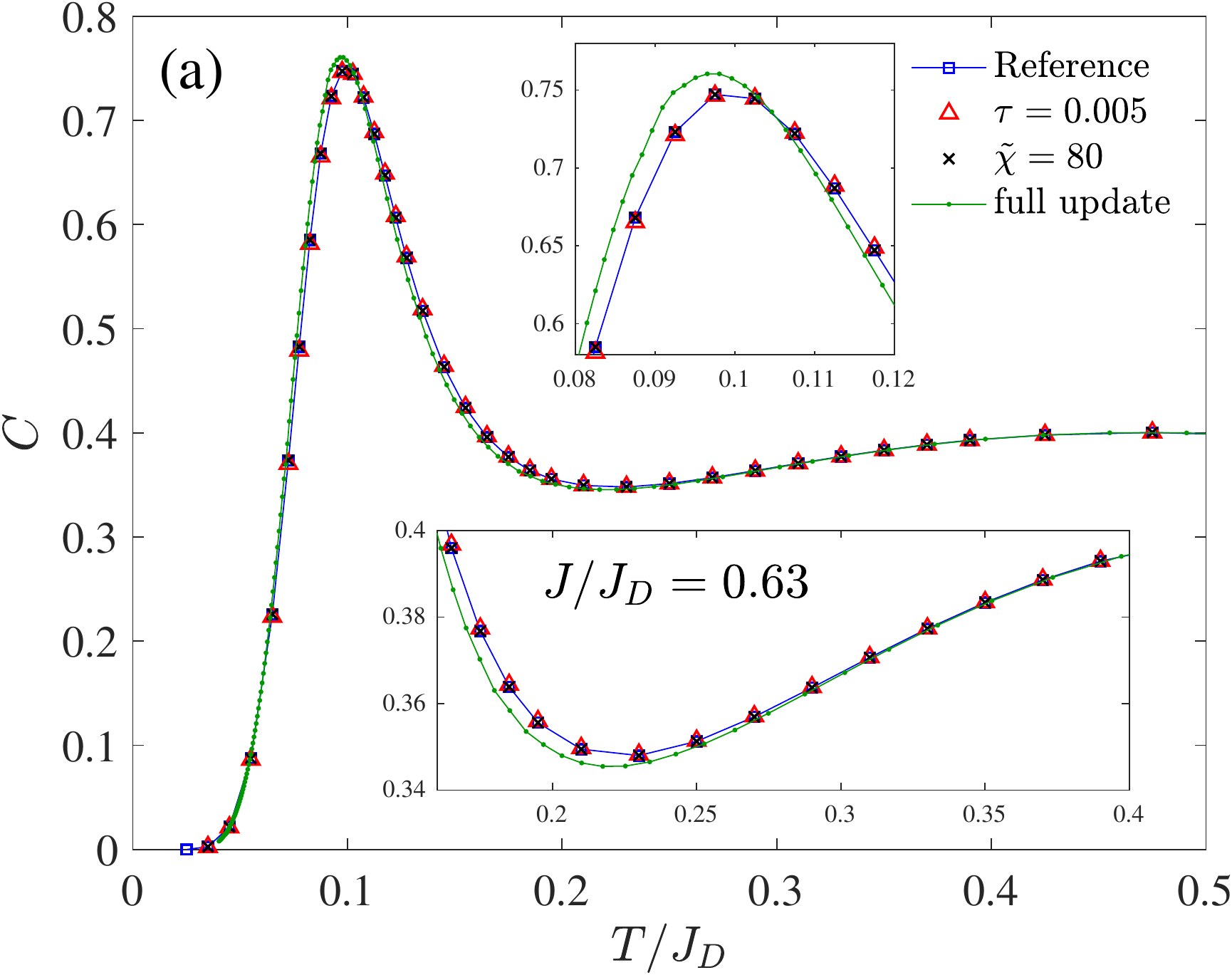}
    \includegraphics[width=1\columnwidth]{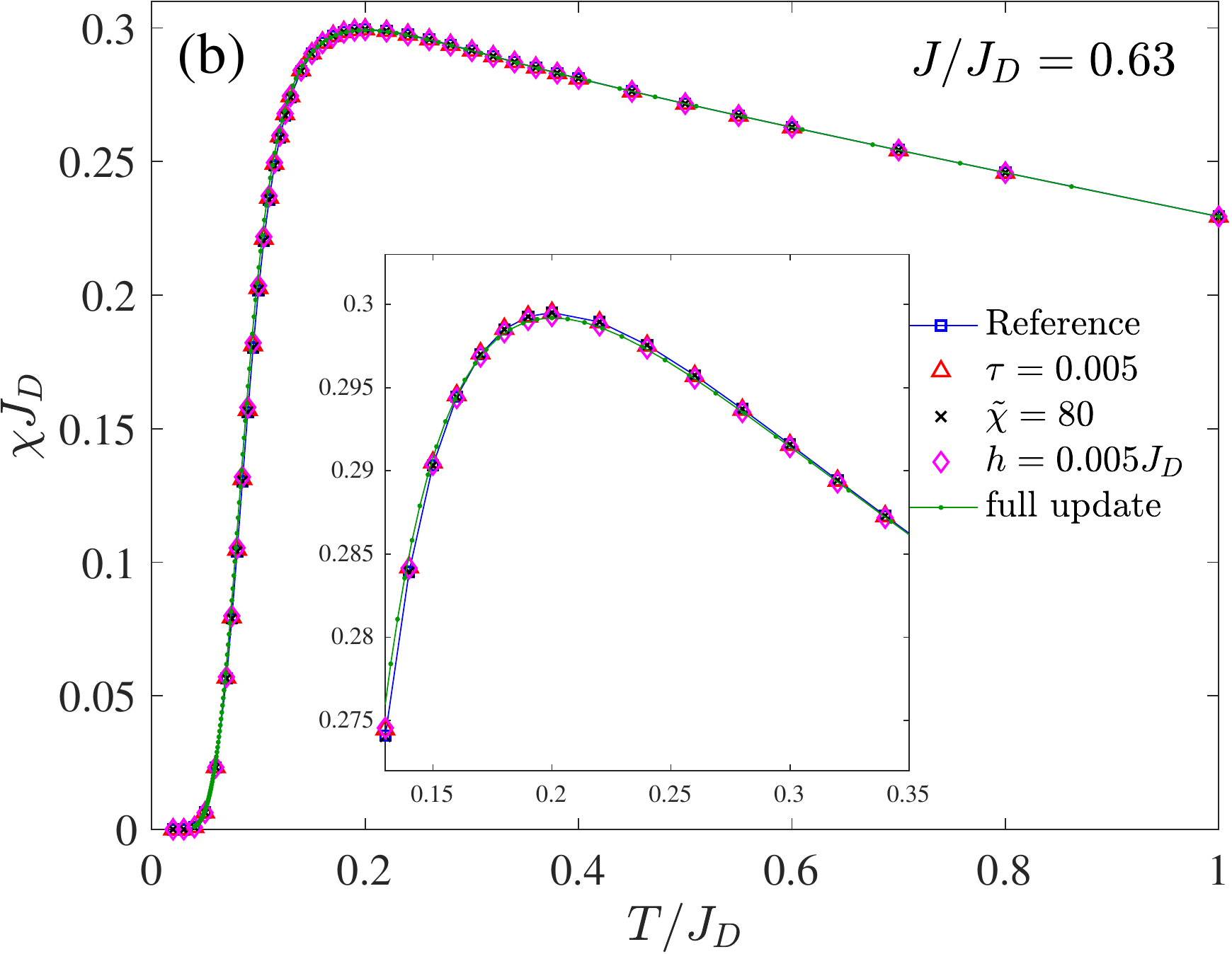}
    \caption{Comparison of $D = 14$ iPEPS data for (a) the magnetic specific 
      heat and (b) the magnetic susceptibility for the Shastry-Sutherland 
      model with $J/J_D = 0.63$. The reference line corresponds to a $D = 14$ 
      simple-update calculation with time step $\tau = 0.04$, boundary bond 
      dimension $\tilde\chi = 50$, and in panel (b) a field $h/J_D = 0.01$. 
      For each of the other lines, one parameter has been varied in the 
      calculation with respect to the reference line, as indicated in the 
      legend. The insets magnify the maxima and the minimum.}
    \label{fig:iPEPS1}
  \end{center}
\end{figure}

\begin{figure}[t]
  \begin{center}
    \includegraphics[width=1\columnwidth]{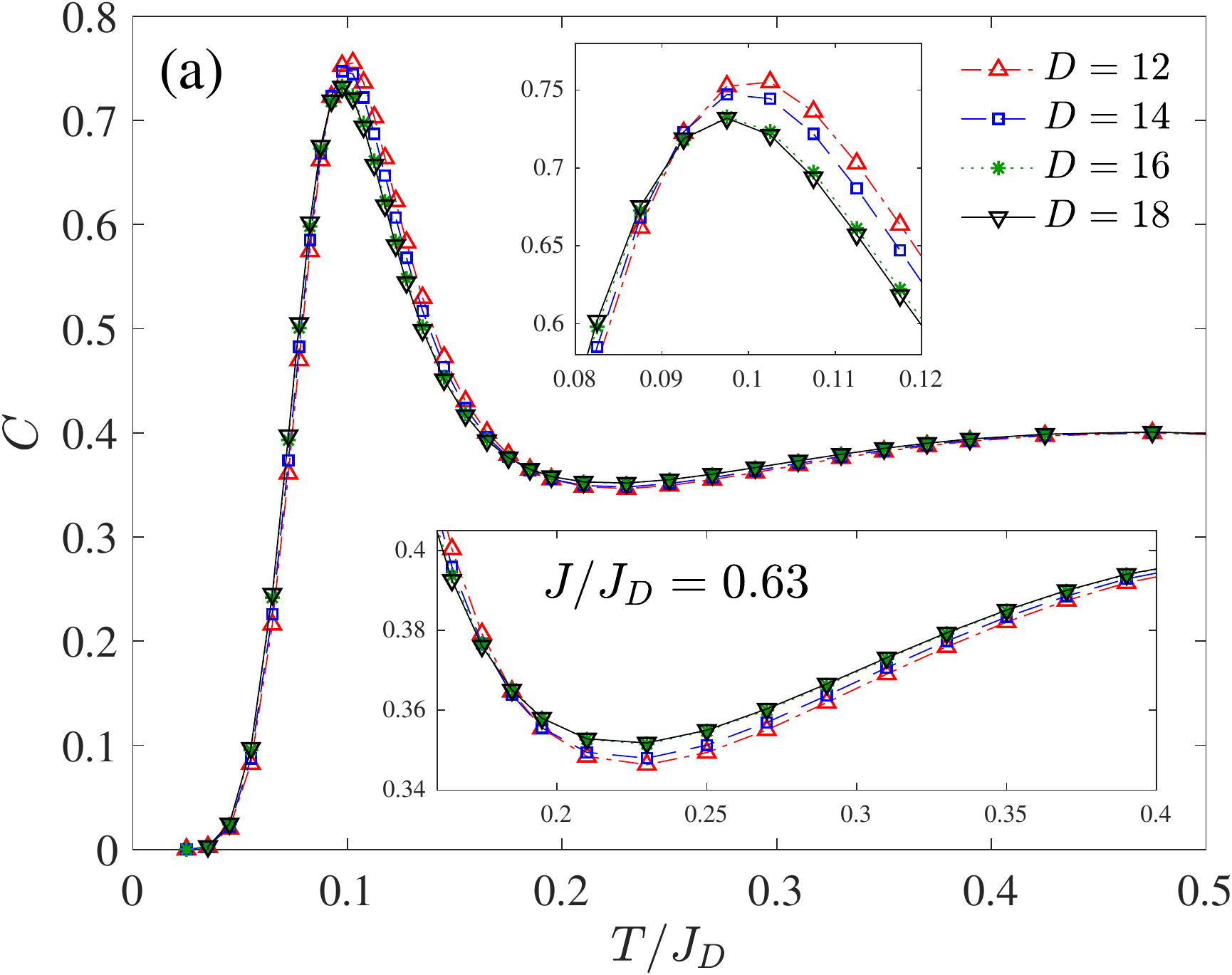}
    \includegraphics[width=1\columnwidth]{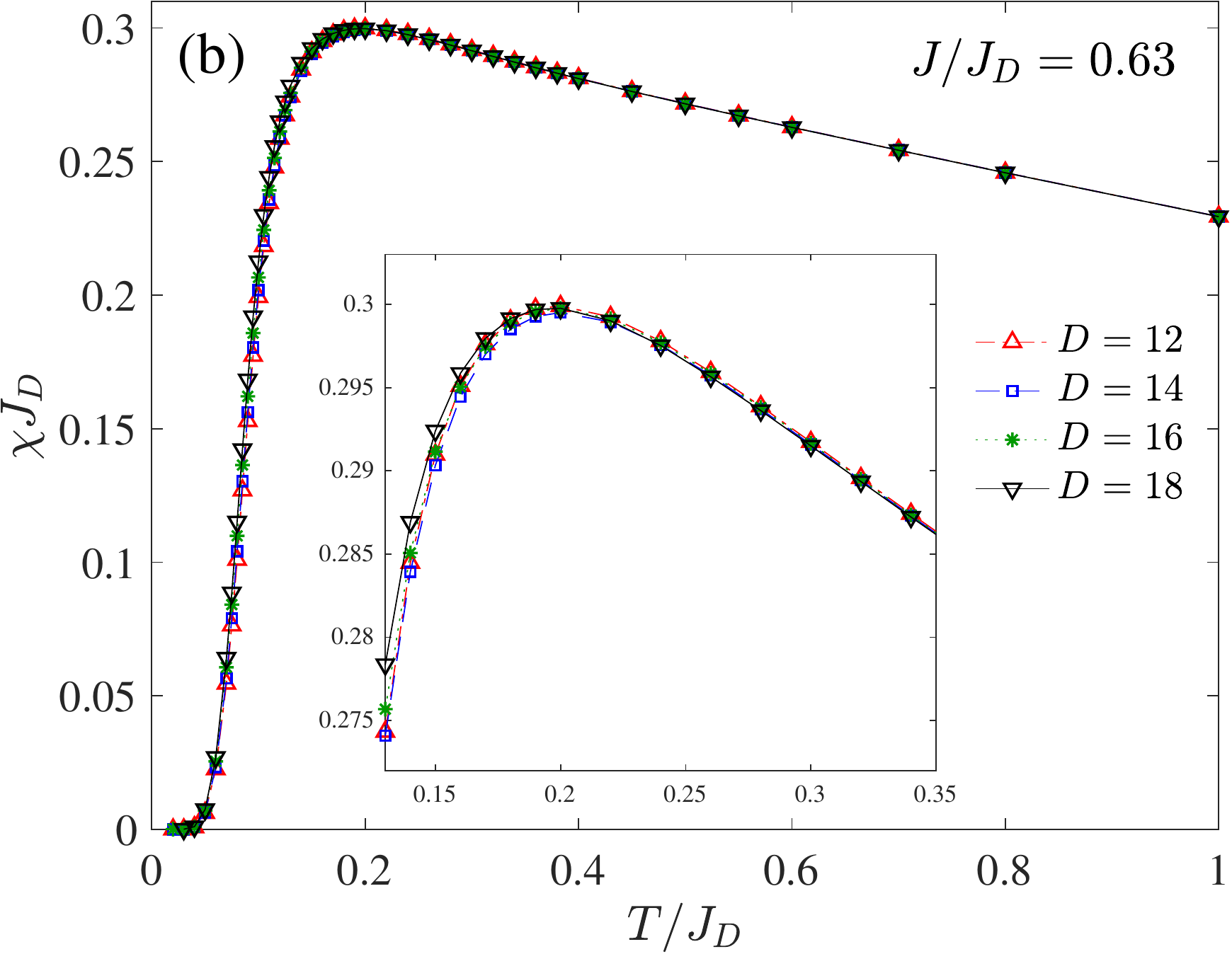}
    \caption{iPEPS results for (a) the magnetic specific heat and (b) the 
      magnetic susceptibility of the Shastry-Sutherland model with $J/J_D
       = 0.63$, obtained by the simple-update method for different values 
      of the bond dimension, $D$. The insets magnify the maxima and the 
      minimum.}
    \label{fig:iPEPS2}
  \end{center}
\end{figure}

Turning to the evaluation of the thermodynamic properties of the 
Shastry-Sutherland model with iPEPS, we obtain the magnetic specific heat 
[Eq.~(\ref{ecdef})] from the numerical derivative of the energy as a function 
of the temperature and the magnetic susceptibility [Eq.~(\ref{echidef})] from 
the numerical derivative of the magnetization with respect to a small external 
magnetic field, $h$. To illustrate the dependence of our results on the 
parameters $\tau$, $D$, and $\tilde \chi$ intrinsic to the iPEPS procedure, 
in Figs.~\ref{fig:iPEPS1} and \ref{fig:iPEPS2} we show example data for the 
coupling ratio $J/J_D = 0.63$.

Our primary results are obtained using a second-order Trotter-Suzuki 
decomposition with a time step $\tau = 0.04$. To demonstrate that this value 
is sufficiently small, meaning that the Trotter discretization error is 
negligible compared to the finite-$D$ errors (below), in Fig.~\ref{fig:iPEPS1} 
we show the comparison with results obtained using a time step $\tau = 0.005$. 
Further, it is easy to show that a relatively small boundary bond dimension, 
such as $\tilde \chi = 50$ for $D = 14$, is large enough for the accurate 
computation of expectation values. As both panels of Fig.~\ref{fig:iPEPS1} 
make clear, results obtained using $\tilde \chi = 80$ are almost identical 
to $\tilde \chi = 50$. To compute the magnetic susceptibility we use a field 
$h/J_D = 0.01$, which is sufficiently small that the discretization error is 
not significant; a comparison with $h/J_D = 0.005$ is shown in 
Fig.~\ref{fig:iPEPS1}(b).

As noted above, the key figure of merit in iPEPS calculations of physical 
quantities is $D$. Because the ``appropriate'' value of $D$ to capture the 
physics of a system varies widely with the problem at hand, we will focus 
throughout our analysis of the Shastry-Sutherland model on benchmarking our 
results by comparing $D$ values. To begin the investigation of $D$-dependence, 
and finite-$D$ convergence, in Fig.~\ref{fig:iPEPS2} we present iPEPS data for 
$J/J_D = 0.63$, obtained by the simple-update method for a range of different 
bond dimensions. We find in Fig.~\ref{fig:iPEPS2}(a) that the position of the 
specific-heat peak shifts slightly ($4\%$) downwards in temperature with 
increasing $D$, while its height is also reduced ($2\%$). However, these 
changes take place largely from $D = 12$ to $D = 16$, whereas changes 
between the $D = 16$ and $D = 18$ data are scarcely discernible on the 
scale of the figure. In the susceptibility [Fig.~\ref{fig:iPEPS2}(b)], 
finite-$D$ effects are clearly very small for $T/J_D > 0.2$; for $0.05 < 
T/J_D < 0.2$ a small increase can be observed with increasing $D$.

Finally, to investigate the difference between results by using the simple- 
and full-update approaches, we return to Fig.~\ref{fig:iPEPS1}. At fixed 
$D = 14$, this difference is relatively small in the specific heat, 
amounting to changes of less than $2\%$ in the position and height of the 
peak [inset, Fig.~\ref{fig:iPEPS1}(a)]. For the susceptibility, differences 
between the two approaches are minimal [Fig.~\ref{fig:iPEPS1}(b)]. 
Because this difference is small in comparison with the finite-$D$ effects 
shown in Fig.~\ref{fig:iPEPS2}, we focus henceforth on the simple-update 
approach, which allows us to reach bond dimensions as large as $D = 20$ in 
the present problem, rather than the full-update approach, where $D = 14$ 
represents a practical upper limit. 

\begin{figure}[t]
  \includegraphics[width=\columnwidth]{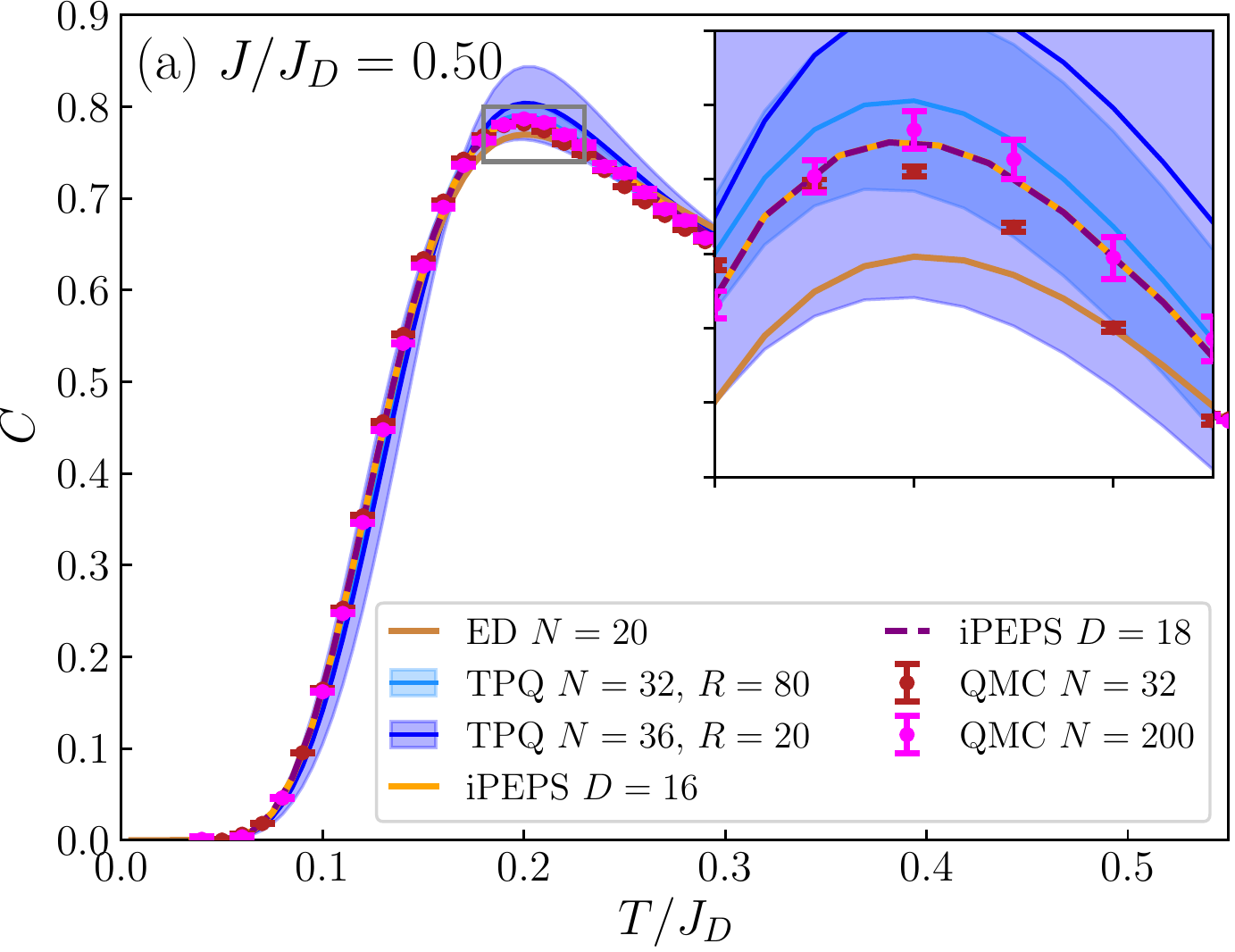}
  \includegraphics[width=\columnwidth]{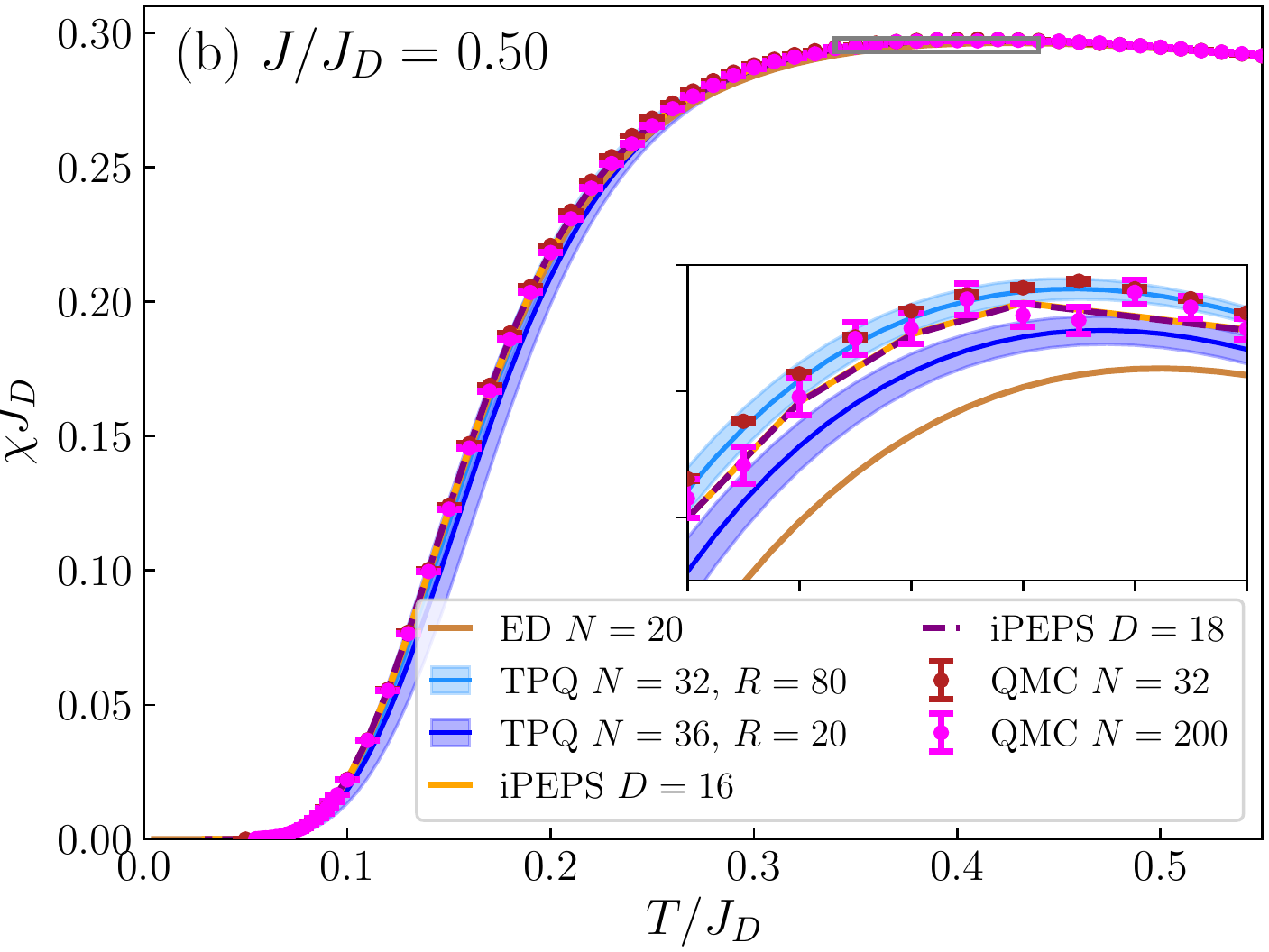}
  \caption{(a) Magnetic specific heat of the Shastry-Sutherland model at 
    $J/J_D = 0.50$, computed by TPQ with $N = 32$ and 36 and by iPEPS with 
    $D = 16$ and 18. Shown for comparison are ED results with $N = 20$ and 
    QMC results with $N = 32$ and 200. (b) Corresponding magnetic 
    susceptibility.}
  \label{fig:sspd05}
\end{figure}

\section{Thermodynamics of the Shastry-Sutherland model with
  $0.60 \le J/J_D \le 0.66$}
\label{tresults}

We begin the systematic presentation and comparison of our thermodynamic 
results by making contact with the previous state of the art, as shown in 
Ref.~\cite{Wessel2018}. The analysis of these authors, which included ED, 
QMC, and high-temperature series expansions (HTSEs), terminated at the 
coupling ratio $J/J_D = 0.60$, where all three methods were beyond their 
limits. However, it was shown that QMC remains a highly accurate technique, 
with only minimal sign problems and thus applicable with large system sizes, 
for $J/J_D < 0.526$. Thus we first show, in Fig.~\ref{fig:sspd05}, the 
magnetic specific heat and susceptibility for $J/J_D = 0.50$ computed by 
TPQ for system sizes $N = 32$ and 36 and by iPEPS with $D = 16$ and 18. 
By comparison with the large-system QMC benchmark ($N = 200$ sites), we 
observe that our iPEPS data, arguably a product of the newest and least 
well-characterized technique, agree precisely for both bond dimensions 
shown. 

For ED-based methods, it was known already \cite{Wessel2018} that 
clusters of size $N = 20$ give an adequate account of the thermodynamics 
at this coupling ratio, with the exception of the very peak in $C(T)$ 
[Fig.~\ref{fig:sspd05}(a)]. Nevertheless, our TPQ results contain two 
surprises, in the form of the rather large errors and the significant 
difference between results for the two cluster sizes. Regarding how 
accurately TPQ reproduces both $C(T)$ and $\chi(T)$ at temperatures around 
the $C(T)$ peak, we note that our $N = 32$ results do indeed lie close to 
the QMC benchmark for this system size, with the discrepancy being well 
within the error tube. Quite generally, the origin of these somewhat large 
errors lies in the very low density of low-lying states in the spectrum, as 
pointed out in Ref.~\cite{Sugiura2013}, and this is certainly the case in the 
Shastry-Sutherland model when the gap is as large as its value at $J/J_D = 
0.5$ (more details may be found in Sec.~\ref{bla}A). We will show that our 
TPQ results become progressively more accurate as the system approaches the 
QPT, which reduces the gap, until coupling ratios very close to the transition. 
With regard to the effects of the shape or geometry of the cluster, regrettably 
the method allows little control beyond the absolute value of $N$. 

\begin{figure}[t]
  \includegraphics[width=\columnwidth]{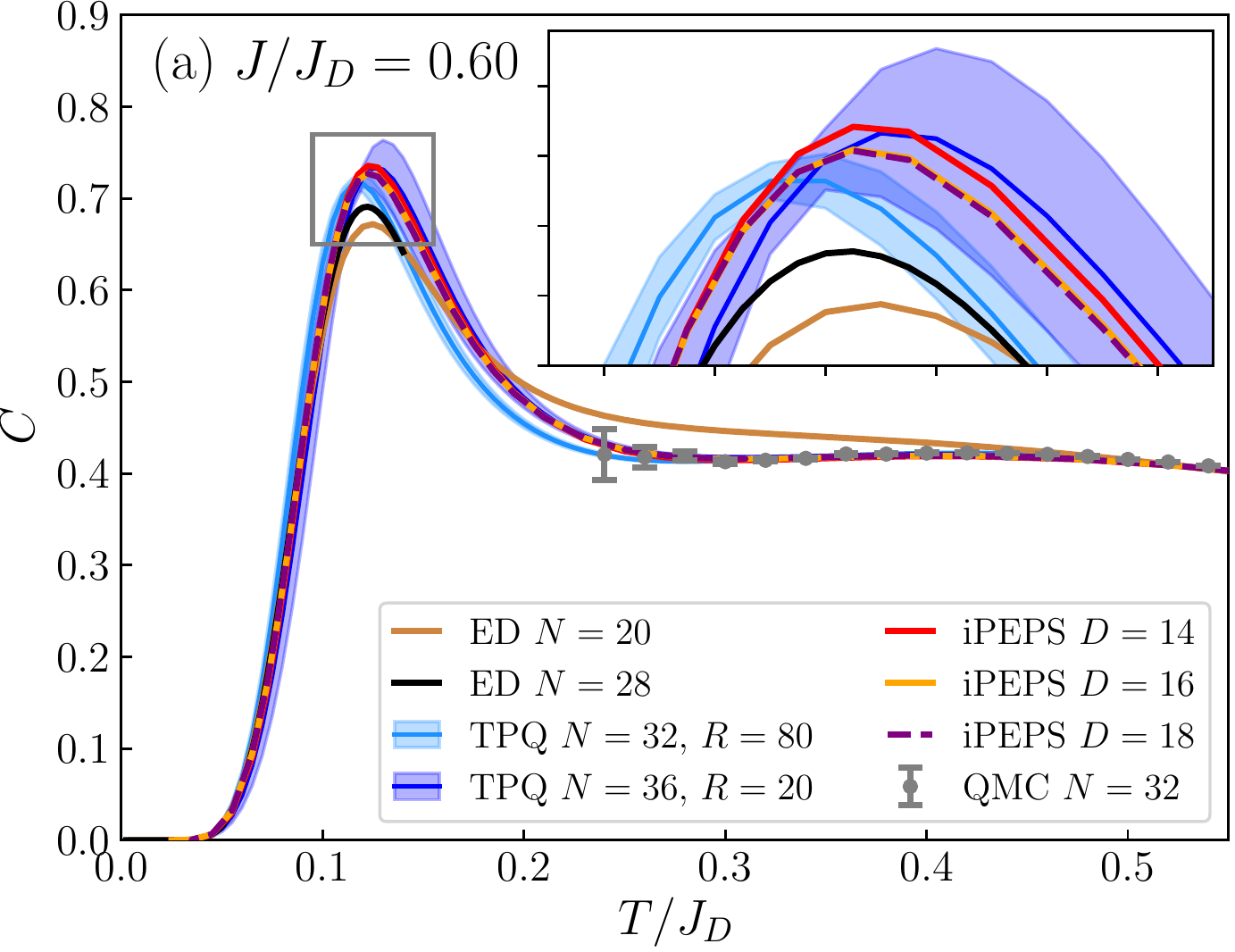}
  \includegraphics[width=\columnwidth]{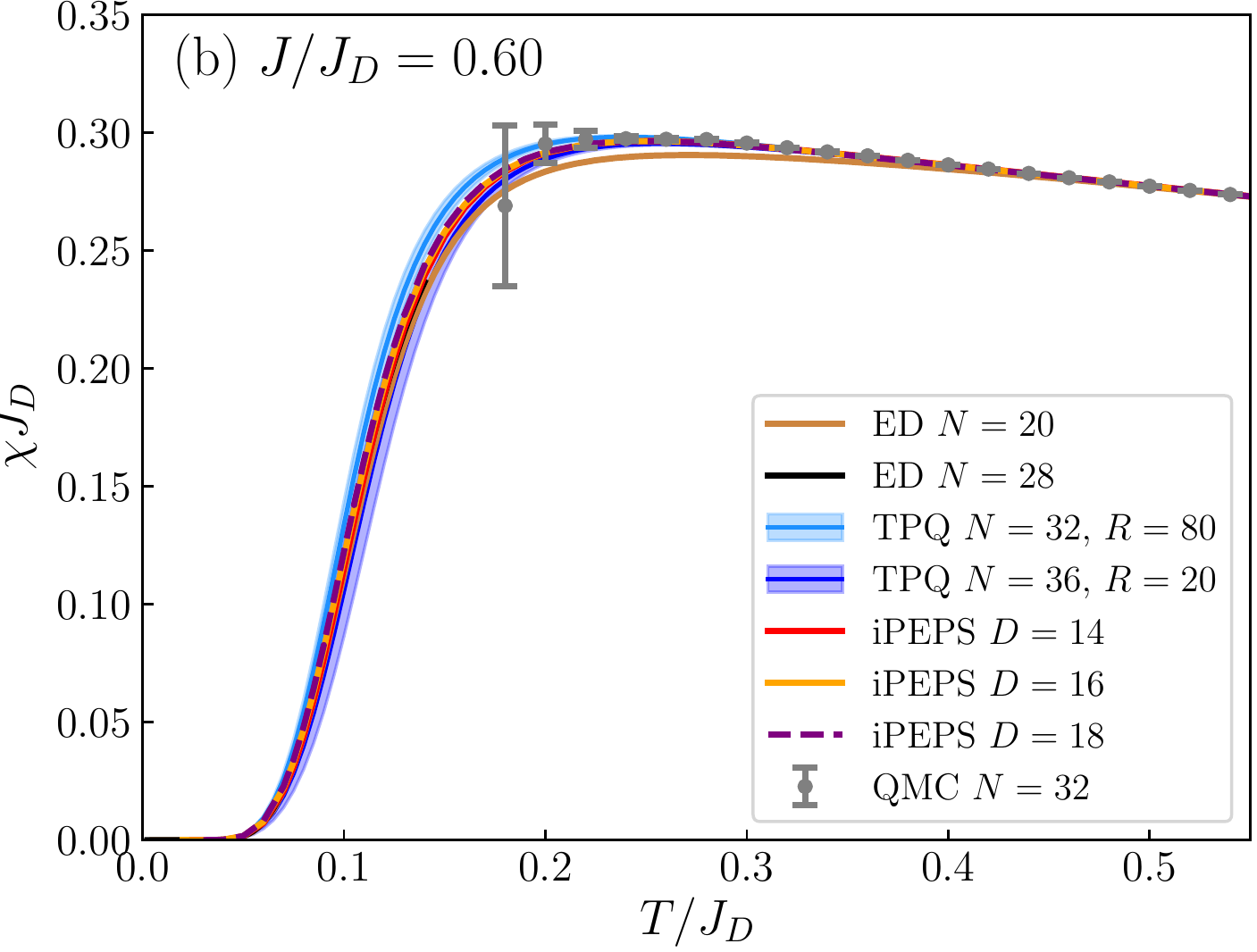}
  \caption{(a) Magnetic specific heat of the Shastry-Sutherland model at 
    $J/J_D = 0.60$, computed by TPQ with $N = 32$ and 36 and by iPEPS with 
    $D = 14$, 16, and 18. Shown for comparison are ED results with $N = 20$ 
    and with $N = 28$ at temperatures $T/J_D \le 0.14$, as well as QMC results 
    with $N = 32$ from Ref.~\cite{Wessel2018}. (b) Corresponding magnetic 
    susceptibility.}
  \label{fig:sspd06}
\end{figure}

In Fig.~\ref{fig:sspd06} we move into the previously intractable parameter 
regime by showing the specific heat and susceptibility at $J/J_D = 0.60$ 
computed by TPQ for system sizes $N = 32$ and 36 and by iPEPS with $D = 14$, 
16, and 18. Focusing first on the specific heat [Fig.~\ref{fig:sspd06}(a)], 
we still observe a single peak at low temperatures ($T/J_D \simeq 0.12$), but 
followed by a very flat plateau extending to $T/J_D \simeq 0.45$. It is clear 
that our TPQ results, augmented by an ED calculation with $N = 28$ 
reaching temperatures just above the peak, exhibit finite-size effects 
in this region. Both the position and the sharpness of the peak show a 
significant evolution as the system size is augmented, although from $N = 32$ 
to 36 the difference is within the error bars of the method. As expected from 
Sec.~\ref{tipeps}, the three iPEPS bond dimensions show good convergence, in 
fact to a peak shape exactly between the two TPQ estimates. Outside the peak 
region, all methods and sizes agree extremely well, and also converge at high 
temperatures to the values obtained by QMC, which returns results of 
acceptably low statistical uncertainty at $T/J_D \gtrsim 0.25$ for this 
coupling ratio \cite{Wessel2018}. 

The corresponding susceptibility [Fig.~\ref{fig:sspd06}(b)] is less 
sensitive to differences in either system size or bond dimension. Indeed, 
only an ED calculation with $N = 20$ is not capable of capturing the maximum 
of $\chi(T)$ for this coupling ratio. However, the two TPQ results do differ 
concerning the exact location in temperature of the rapid rise in $\chi(T)$, 
and the fully convergent iPEPS results appear to offer the benchmark. All 
of our calculations agree closely on the location of the broad maximum. 
We comment that the characteristic temperature of $\chi(T)$ need not match 
that of $C(T)$, given that the former has contributions only from magnetic 
states but the latter includes the contributions of singlets. However, 
at this coupling ratio, where the one-triplon excitations of the system lie 
below all two-triplon bound states \cite{Wessel2018}, the peak in $C(T)$ and 
the rise of $\chi(T)$ do indeed coincide. We discuss the excitation spectrum 
of the model in detail in Sec.~\ref{bla}A.

\begin{figure}[t]
  \includegraphics[width=\columnwidth]{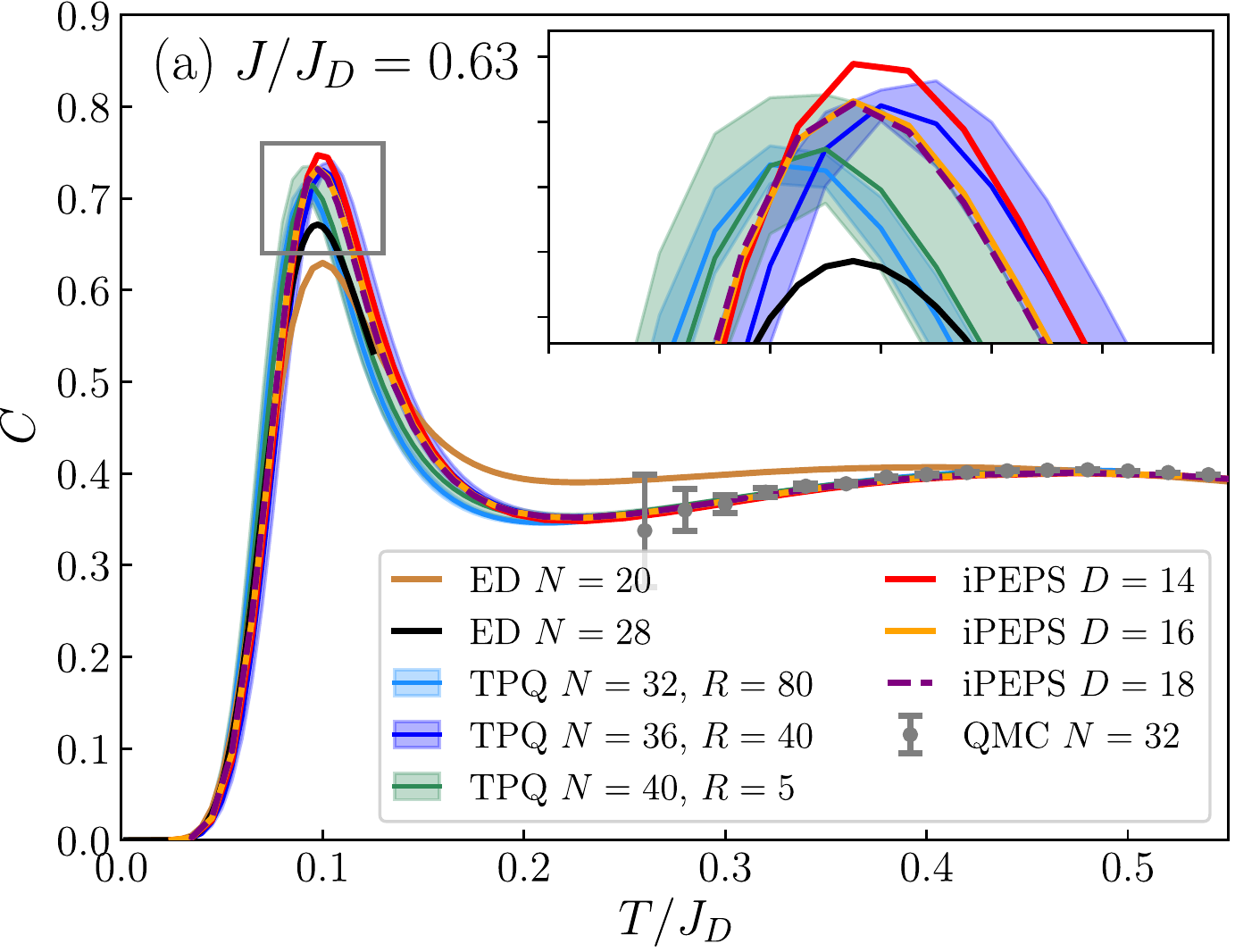}
  \includegraphics[width=\columnwidth]{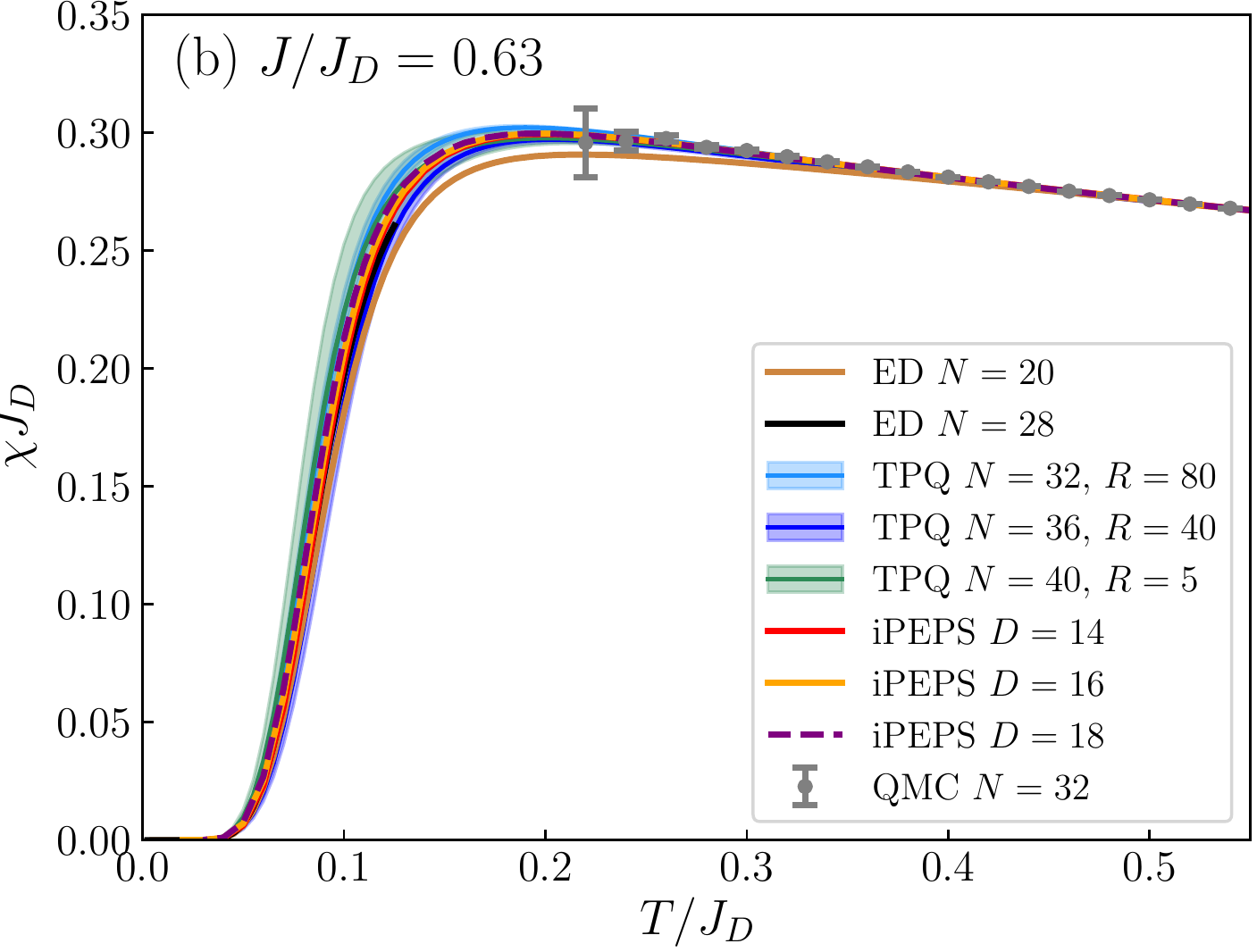}
  \caption{(a) Magnetic specific heat of the Shastry-Sutherland model at 
    $J/J_D = 0.63$, computed by TPQ with $N = 32$, 36, and 40 and by iPEPS with 
    $D = 14$, 16, and 18. Shown for comparison are ED results with $N = 20$ and 
    with $N = 28$ at temperatures $T/J_D \le 0.125$, as well as QMC results 
    with $N = 32$. (b) Corresponding magnetic susceptibility.}
  \label{fig:sspd063}
\end{figure}

As noted in Sec.~\ref{sintro}, the coupling ratio $J/J_D = 0.63$ has special 
significance in the Shastry-Sutherland model because of its proposed connection 
to the material SrCu$_2$(BO$_3$)$_2$, and for this reason we have already used 
it for benchmarking purposes in Secs.~\ref{ttpq} and \ref{tipeps}. In 
Fig.~\ref{fig:sspd063} we compare our TPQ and iPEPS results for $C(T)$ and 
$\chi(T)$. Independent of either method, we observe that the specific-heat 
peak has become sharper, although not any taller, and that the drop on its 
high-$T$ side has turned into a true minimum, before $C(T)$ recovers to a 
broad maximum around $T/J_D = 0.45$. The position of the sharp peak has moved 
down to $T/J_D \simeq 0.10$, which is a remarkably large shift for such a small 
change in coupling ratio. Although the corresponding features are far less 
pronounced in $\chi(T)$, it it clear that onset is considerably steeper 
than at $J/J_D = 0.60$ and that the temperature scales both for it and for 
the flat maximum are lower. 

Focusing on the details of $C(T)$ [Fig.~\ref{fig:sspd063}(a)], once again our 
TPQ results for $N = 32$ and 36, particularly in combination with $N = 28$ ED, 
show nontrivial changes around the peak, both in its position and its height. 
However, the $N = 40$ data appear to offer a systematic interpolation of both 
quantities. In our iPEPS results, $D = 14$ no longer appears representative 
of the large-$D$ limit at this coupling ratio, but finite-$D$ effects seem to 
become very small at higher $D$. Qualitatively, our optimal iPEPS peak appears 
to be very close to the form at which one might expect the TPQ results to 
converge with system size, and we will quantify this statement below. In 
$\chi(T)$ [Fig.~\ref{fig:sspd063}(b)], the two methods achieve full convergence 
for all $N$ and $D$, despite the lowering effective temperature scales, and we 
draw attention to the fact that previous ED and QMC studies at this coupling
ratio could not capture this low-temperature behavior with any accuracy; we 
revisit this point in the context of SrCu$_2$(BO$_3$)$_2$ in Sec.~\ref{bla}C.

\begin{figure}[t]
  \includegraphics[width=\columnwidth]{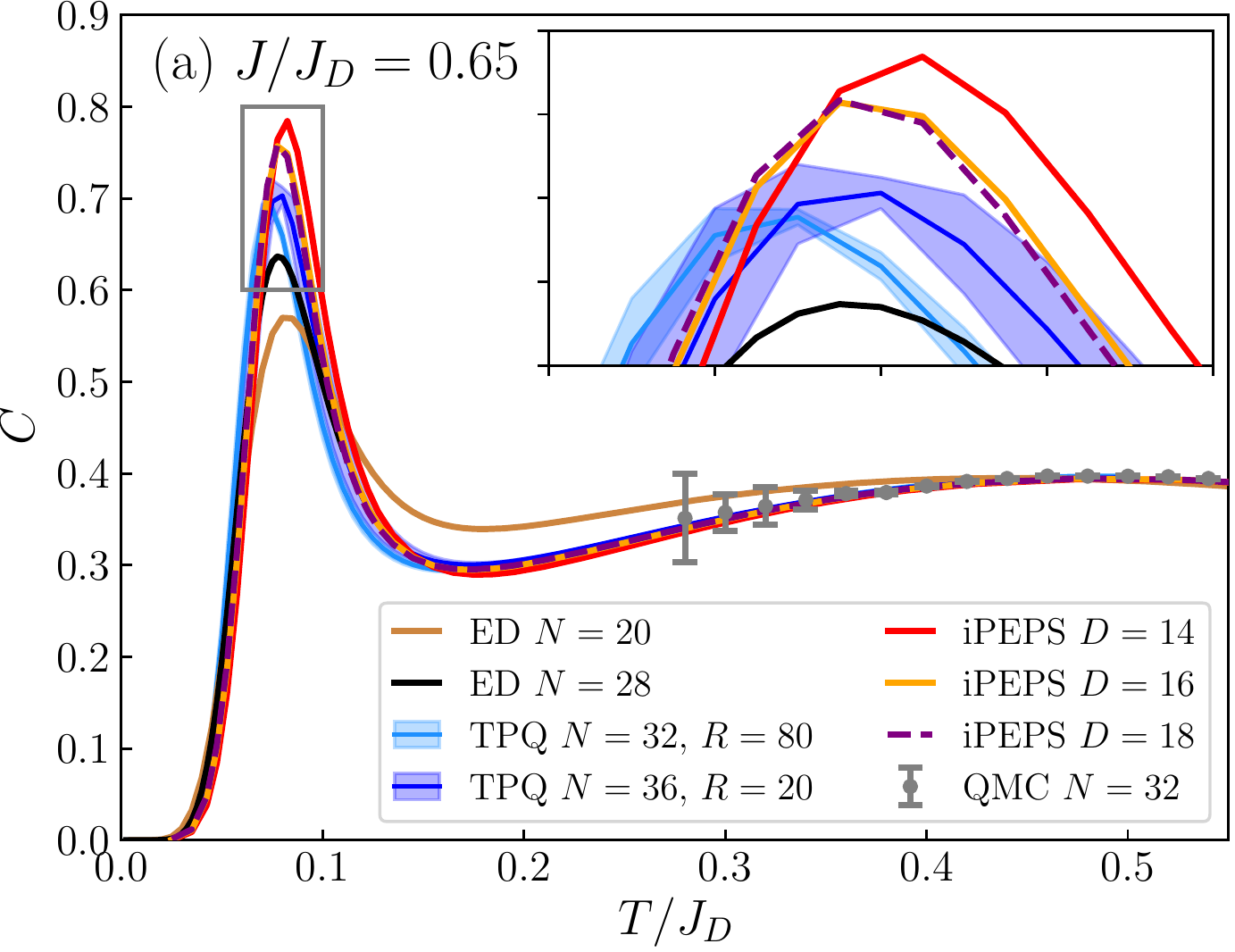}
  \includegraphics[width=\columnwidth]{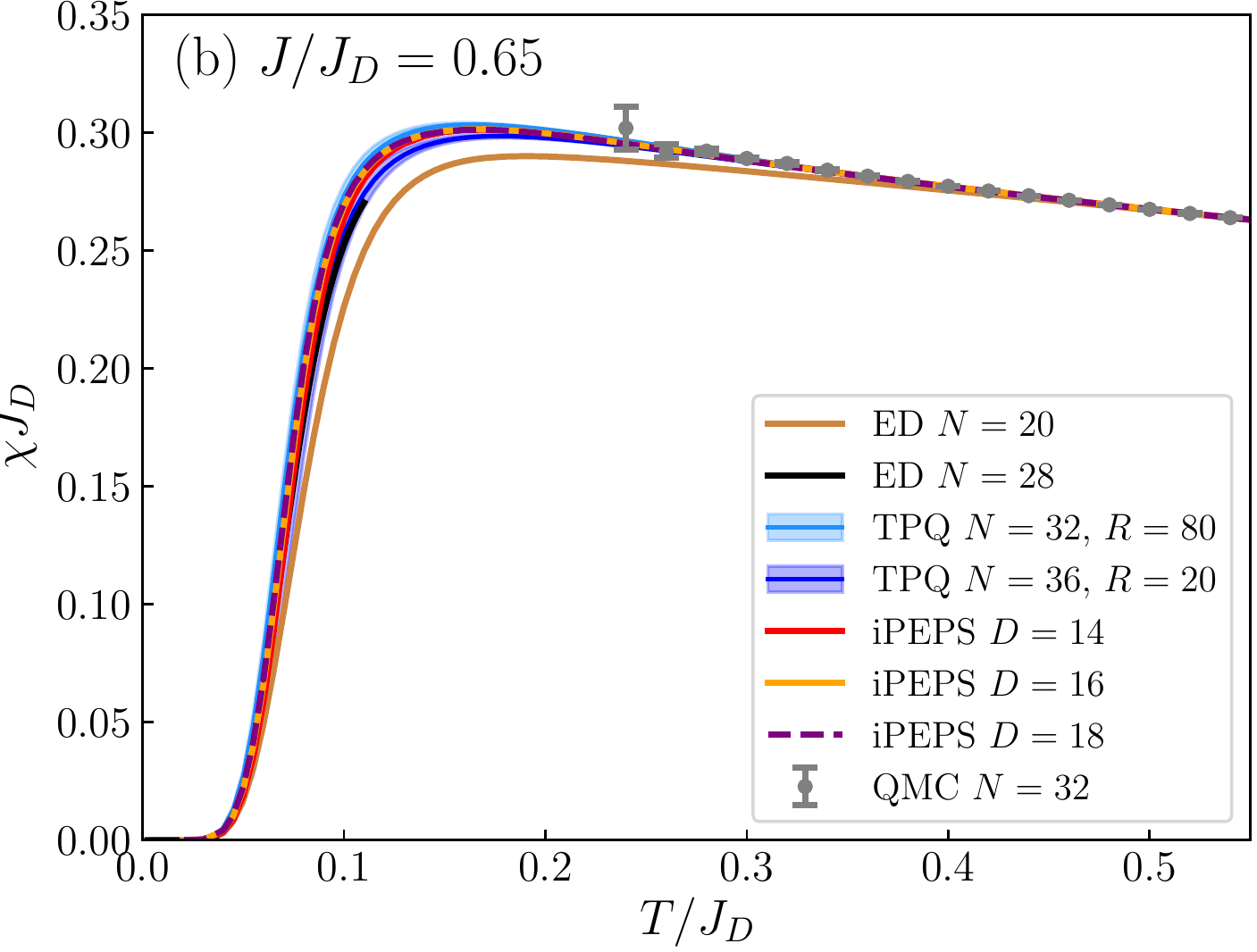}
  \caption{(a) Magnetic specific heat of the Shastry-Sutherland model at 
    $J/J_D = 0.65$, computed by TPQ with $N = 32$ and 36 and by iPEPS with 
    $D = 14$, 16, and 18. Shown for comparison are ED results with $N = 20$
    and with $N = 28$ at temperatures $T/J_D \le 0.11$, as well as QMC 
    results with $N = 32$. (b) Corresponding magnetic susceptibility.} 
  \label{fig:sspd065}
\end{figure}

Proceeding yet closer to the QPT, Fig.~\ref{fig:sspd065} shows the specific 
heat and susceptibility at the coupling ratio $J/J_D = 0.65$. Clearly the 
``peak-dip-hump'' shape of $C(T)$ has become significantly more pronounced 
in all respects than at $J/J_D = 0.63$; the peak is sharper, the dip is deeper, 
and the separation of the peak and the broad hump has increased, with the peak 
moving below $T/J_D = 0.09$ and the high-$T$ maximum centered around $T/J_D = 
0.5$. Once again the progression from our ED data at $N = 28$ to TPQ with 
$N = 32$ and then 36 shows an increasing ability to capture the tip of the 
peak, but it cannot be said that convergence has been reached [inset, 
Fig.~\ref{fig:sspd065}(a)]. As at $J/J_D = 0.63$, our iPEPS data for 
$D = 16$ and 18 are in essence indistinguishable and do suggest the peak 
to which the TPQ results are converging. At all other parts of the $C(T)$ 
curve, convergence is complete by all methods. In the corresponding 
susceptibility, again the ED and TPQ size sequence provides a systematic 
convergence towards the ``squareness'' of the turnover from rapid rise to 
broad maximum, while the higher iPEPS bond dimensions have converged to 
the shape of this feature. 

\begin{figure}[t]
  \includegraphics[width=\columnwidth]{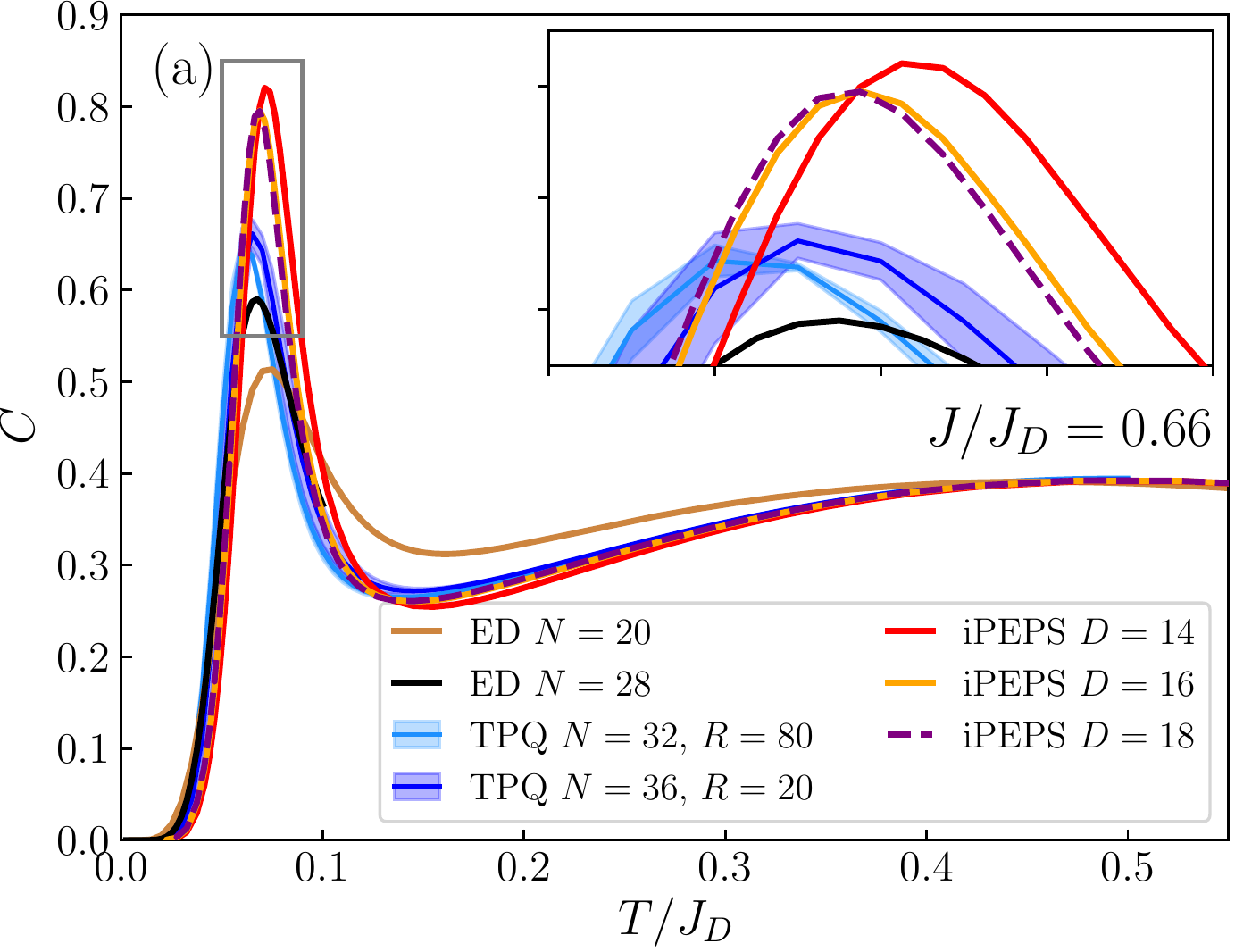}
  \includegraphics[width=\columnwidth]{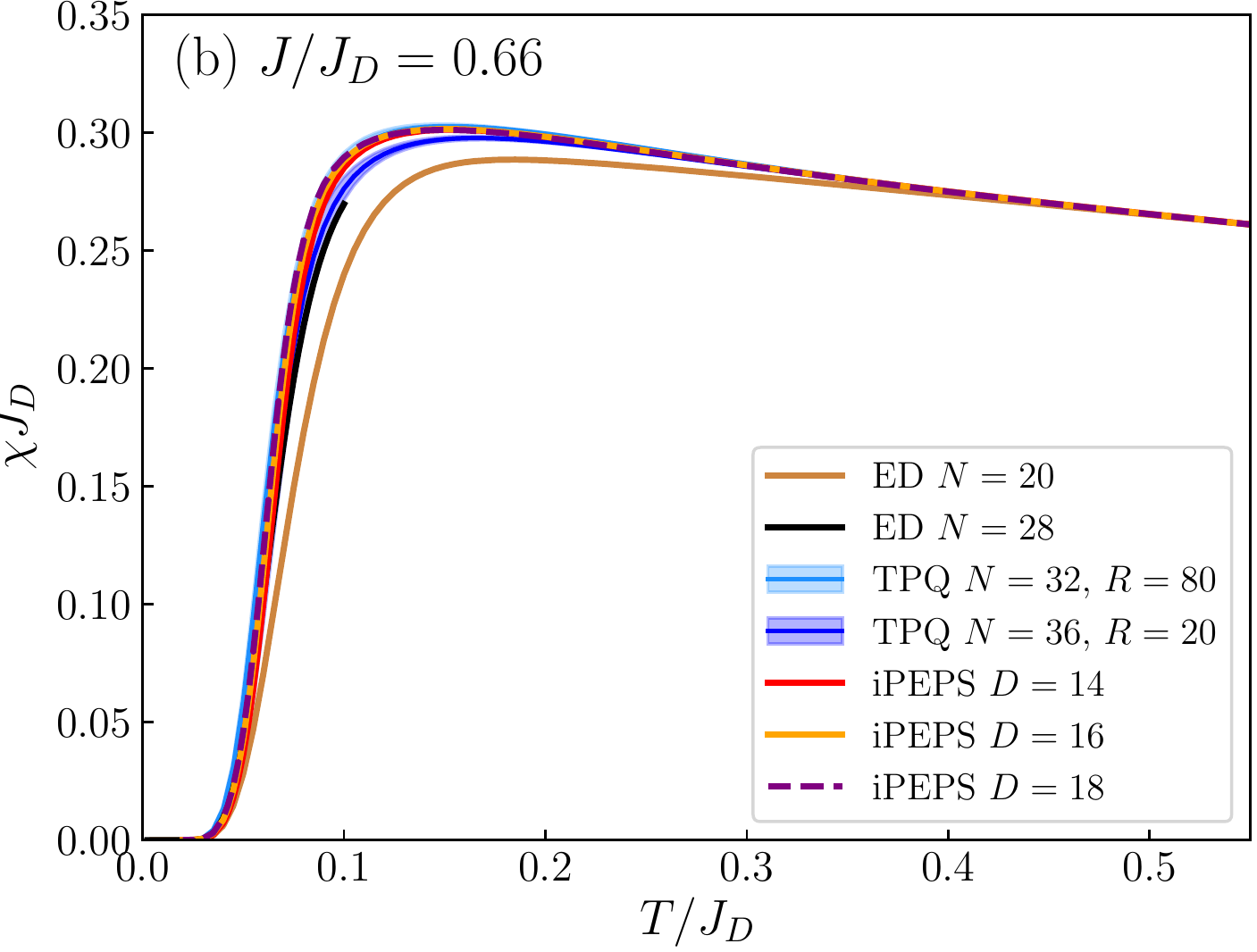}
  \caption{(a) Magnetic specific heat of the Shastry-Sutherland model at 
    $J/J_D = 0.66$, computed by TPQ with $N = 32$ and 36 and by iPEPS with 
    $D = 14$, 16, and 18. Shown for comparison are ED results with $N = 20$
    and with $N = 28$ at temperatures $T/J_D \le 0.10$. (b) Corresponding 
    magnetic susceptibility.} 
  \label{fig:sspd066}
\end{figure}

We take the coupling ratio $J/J_D = 0.66$, shown in Fig.~\ref{fig:sspd066}, 
as the upper limit to the validity of comparing our methods. Physically, it 
is obvious that the peak-dip-hump shape of $C(T)$ becomes yet more pronounced, 
that the dip is deeper, and that the peak is even lower-lying and sharper than 
at $J/J_D = 0.65$ [Fig.~\ref{fig:sspd066}(a)]. Numerically, it appears that 
both our methods show slower convergence, not only around the peak 
but also in the dip. In more detail, our $N = 32$ and 36 TPQ results, and our 
$D = 16$ and 18 iPEPS results, do both offer internally consistent peak shapes. 
However, it is no longer clear that TPQ and iPEPS might extrapolate to the 
same limit, although this could be because the effective location of the QPT
differs for the two methods due to finite-$D$ and -$N$ corrections, implying
closer proximity to a competing phase in one than in the other. On the 
overall shape of $C(T)$, we comment that the hump can be understood as the 
energy scale of local spin-flipping processes on the scale of $J_D$, while the 
low-$T$ peak is the focus of our discussions in Sec.~\ref{bla}A. Concluding 
with the susceptibility at $J/J_D = 0.66$ [Fig.~\ref{fig:sspd066}(b)], again 
our iPEPS results offer a consistent picture of the very steep onset and 
square maximum, while our TPQ results suggest better agreement with iPEPS 
at $N = 32$ than at $N = 36$. 

The results in Figs.~\ref{fig:sspd06} to \ref{fig:sspd066} capture the 
thermodynamic properties of the Shastry-Sutherland model with a degree 
of precision that was entirely unattainable by all previous techniques. 
It is safe to say that no prior numerical studies had even identified 
the peak-dip-hump form of the purely magnetic specific heat with any 
reliability, let alone given an indication of how narrow the low-$T$ 
peak becomes in the regime near the QPT. While the square shape of 
$\chi(T)$ has been known since the early measurements on SrCu$_2$(BO$_3$)$_2$ 
\cite{Kageyama1999,Kageyama1999b}, again no numerical approaches had reproduced 
this form. We discuss both the physics underlying these features and the 
experimental comparison in Sec.~\ref{bla}. 

The thermodynamic properties of the Shastry-Sutherland model 
nevertheless set an extremely challenging problem in the regime close 
to the QPT, even in the dimer-product phase where the ground state is 
exact and no finite-temperature transition is expected. It is clear in 
Figs.~\ref{fig:sspd06} to \ref{fig:sspd066} that neither our TPQ nor our 
iPEPS calculations can be judged to be fully convergent, and thus accurate, 
as $J/J_D$ approaches the presumed critical value of 0.675. Because both 
are limited by their control parameters, $N$ for TPQ and $D$ for iPEPS, 
an optimal reproduction of the exact thermodynamics would be obtained by 
extrapolating both sets of results to infinite $N$ or $D$. 

\begin{figure}[t]
  \includegraphics[width=\columnwidth]{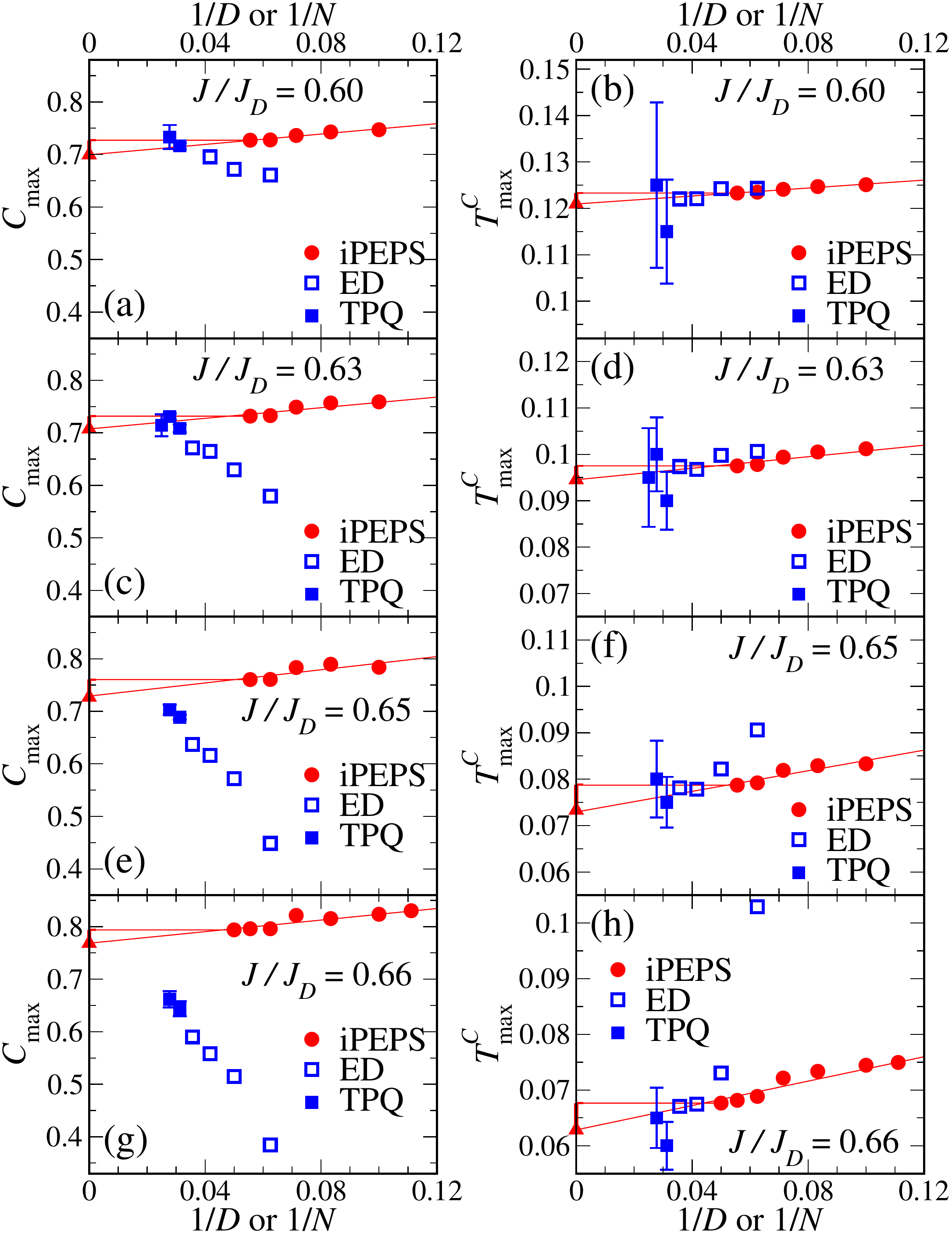}
  \caption{Characteristic position and height of the low-$T$ peak in the 
    magnetic specific heat of the Shastry-Sutherland model, shown for our ED 
    and TPQ results as functions of $1/N$ and for our iPEPS results as 
    functions of $1/D$. (a,b) $J/J_D = 0.60$. (c,d) $J/J_D = 0.63$. (e,f) $J/J_D
     = 0.65$. (g,h) $J/J_D = 0.66$.} 
  \label{fig:ndc}
\end{figure}

By inspection of the full peak shape in $C(T)$, our TPQ results can only be 
said to have reached convergence at $N \le 40$ for $J/J_D \le 0.63$. Similarly, 
it is not clear that our iPEPS results for the largest available $D$ values 
(we have performed calculations with $D = 18$ at all temperatures and with 
$D = 20$ for temperatures around the peak) can be said to have converged to 
the shape of the peak at $J/J_D = 0.66$. For extrapolation purposes, we first 
extract the primary characteristic features of this low-$T$ peak, namely its 
position and height. In Fig.~\ref{fig:ndc} we show the convergence of both 
quantities as functions respectively of $1/N$ in our ED and TPQ calculations 
and of $1/D$ in our iPEPS calculations, for each of the coupling ratios 
$J/J_D = 0.60$, 0.63, 0.65, and 0.66.

Focusing first on our iPEPS results, it is evident that the values we have 
obtained for both the position and the height of the $C(T)$ peak show an 
excellent linear convergence with $1/D$ for all coupling ratios. Both 
quantities decrease with increasing $D$ and in fact their slopes are 
remarkably insensitive to $J/J_D$. Thus we estimate the putative infinite-$D$ 
limit of the peak characteristics (shown by a filled triangle) by linear 
extrapolation in $1/D$ and we take the difference of this limit from the 
result at the largest computed $D$ value as an estimate of the (one-sided) 
error bar. While all of the iPEPS estimates decrease monotonically with 
increasing $D$, in the large-$D$ limit one does anticipate exponential 
convergence in $D$ rather than linear convergence in $1/D$. Thus we expect 
the true $D = \infty$ limit to lie within the interval indicated by the error 
bar.

Turning to the ED and TPQ results in Fig.~\ref{fig:ndc}, the peak 
heights at all coupling ratios show a quite systematic convergence, from 
below and with changing slopes, towards values fully consistent with our 
iPEPS estimates. By contrast, the finite-size evolution of the peak positions 
is less clear, but it is true to state for every coupling ratio that their 
values for all system sizes are either clustered closely around the 
extrapolated iPEPS values or are trending towards these. 

Thus we may conclude that our TPQ and iPEPS results are consistent and 
convergent for even the most challenging features of the most challenging 
region of the Shastry-Sutherland phase diagram. As such they may be used as 
a reliable statement of the excitation spectrum as a function of the coupling 
ratio, which is the topic of Sec.~\ref{bla}A. We discuss the implications of 
the convergence details in Fig.~\ref{fig:ndc} for future numerical development 
in Sec.~\ref{bla}B. 

\section{Discussion}
\label{bla}

\subsection{Spectrum of the Shastry-Sutherland Model}

In Fig.~\ref{fig:sspdmultiple}(a) we showed our collected results for the 
specific heat of the Shastry-Sutherland model across the range of coupling 
ratios $0.60 \le J/J_D \le 0.66$. As the system approaches the QPT from the 
dimer-product to the plaquette state, one observes the systematic emergence of 
a narrow low-energy peak, followed at higher temperatures by a deepening dip 
before a second broad maximum on the scale of $J_D/2$. The energy scale of 
the concentration of low-lying states contributing to the peak falls steeply 
as the QPT is approached. From the corresponding susceptibilities, shown in 
Fig.~\ref{fig:sspdmultiple}(b), it is clear that the decreasing energy scale 
is largely similar for triplet (and higher spinful) excitations, and hence 
at minimum that it is not a special property of singlets; the physics of the 
approach to the QPT seems to affect all excitations in the same general manner. 

\begin{figure}[t!]
  \includegraphics[width=0.9\columnwidth]{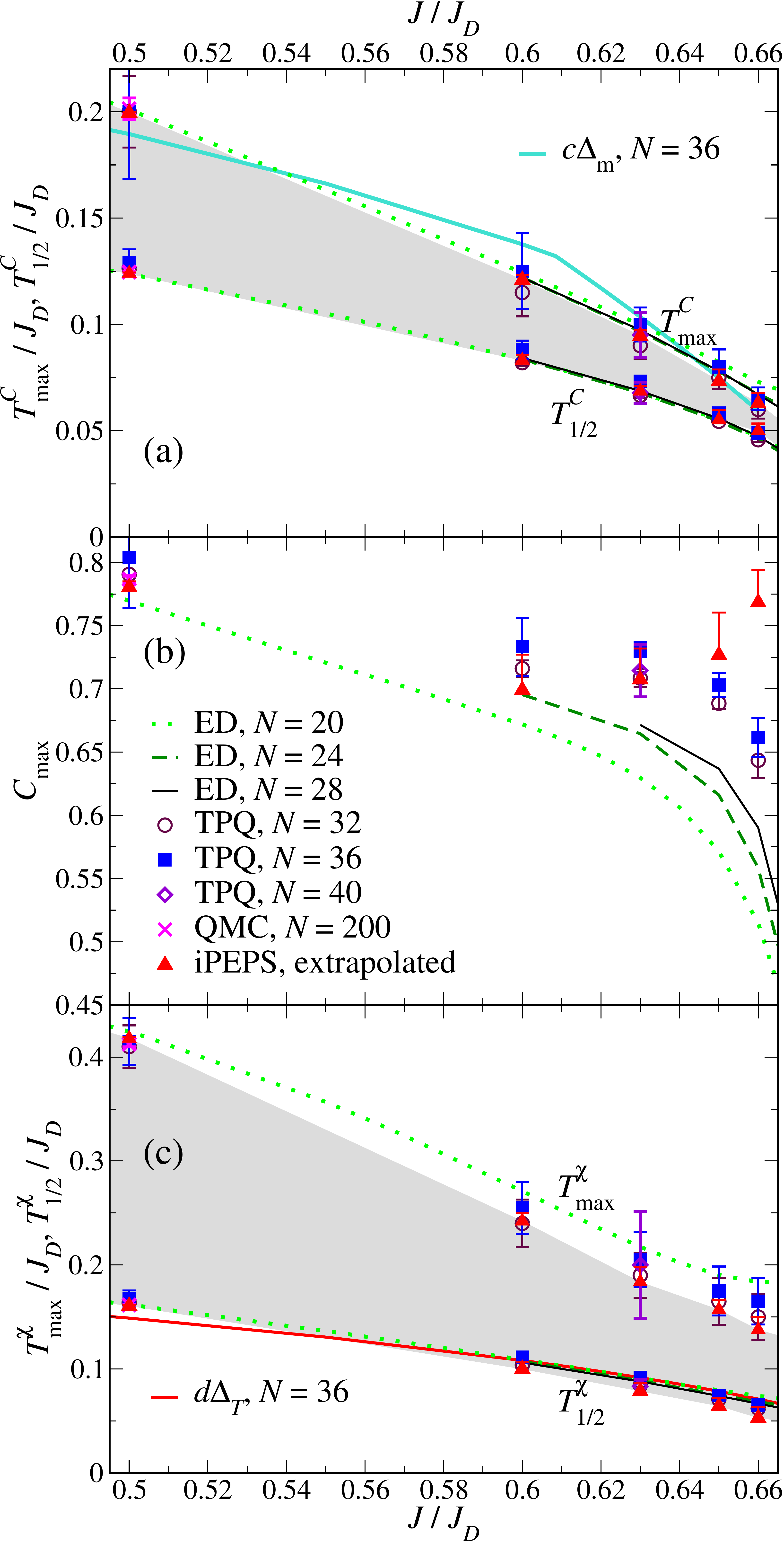}
  \caption{(a) Position, $T_{\max}^C$, and (b) height, $C_{\max}$, of the peak 
    in the specific heat of the Shastry-Sutherland model in the regime of 
    coupling ratios $0.50 \le J/J_D \le 0.66$, as determined from our ED, 
    TPQ, and iPEPS calculations. Also shown in panel (a) is the half-height, 
    $T_{1/2}^C$, on the low side of the peak, from which one may deduce an 
    effective peak width (shaded grey region). iPEPS error-bar conventions 
    are as in Fig.~\protect{\ref{fig:ndc}}. At $J/J_D = 0.50$ we compare in 
    addition with the large-$N$ QMC result. Shown for reference in panel (a)
    is the function $c \, \Delta_{\rm m} (J/J_D)$, where $\Delta_{\rm m} = {\rm 
    min}[\Delta_{S},\Delta_{T}]$, with $\Delta_{S}$ and $\Delta_{T}$ respectively 
    the gaps of the lowest singlet and triplet excitations obtained for a 
    system of size $N = 36$, is the minimal gap in the system; the constant 
    of proportionality is $c = 0.28$. (c) Positions of the peak, $T_{\max}^\chi$, 
    and of the half-height, $T_{1/2}^\chi$, in the susceptibility of the 
    Shastry-Sutherland model in the same range of coupling ratios. Again 
    the shaded region may be considered as an effective width. Shown for 
    reference is the function $d \, \Delta_T$, again obtained for $N = 36$, 
    with constant of proportionality $d = 0.22$.}
  \label{fig:ph}
\end{figure}

Before discussing the nature of the low-lying spectrum, we characterize the 
height and temperature scale of the specific-heat peak, and the temperature 
scale of the susceptibility onset, as the coupling ratio approaches the QPT. 
The peak position, $T_{\max}^C$, shown in Fig.~\ref{fig:ph}(a), clearly falls 
faster than linearly in $J/J_D$, but does not approach zero at the QPT. Based 
on the expectation that the peak position as a function of $J/J_D$ should 
follow the gap, $\Delta_{\rm m}$, of the system, in Fig.~\ref{fig:ph}(a) we show 
this quantity based on ED data from the $N = 36$ cluster~\cite{Wessel2018}, 
and with a multiplicative prefactor of 0.28. The fact that these
falling energy scales do not vanish at the QPT is consistent with the 
first-order nature of this transition \cite{Corboz13_shastry}. The behavior 
of the peak height [Fig.~\ref{fig:ph}(b)] is less evident, in that it falls 
on approach to the QPT in our TPQ calculations but increases in our iPEPS 
ones, the rate of change accelerating with $J/J_D$ in both cases. Given the 
strong finite-size effects visible in our ED and TPQ calculations for $J/J_D
 > 0.63$ [Fig.~\ref{fig:ndc}], here it appears that our iPEPS results are more 
representative of the large-$N$ limit. The final piece of information one 
may wish to extract about the peak is its width, for which we show in addition 
the temperature $T_{1/2}^C$, at which the specific heat attains half its peak 
height [Fig.~\ref{fig:ph}(a)]; because of the dip and hump on the high-$T$ 
side, we focus only on the low-$T$ side. We obtain a characteristic width 
from the quantity $T_{\max}^C - T_{1/2}^C$, shown as the shaded grey region. We 
observe that the half-height and full-height positions, and hence the width, 
scale to a good approximation in the same way with $J/J_D$, separated only 
by constant factors, and conclude that all the peak shapes are actually 
self-similar on the low-temperature side, with only one characteristic 
$J/J_D$-dependent energy scale. 

\begin{figure}[t]
  \centering
  \includegraphics[width=0.97\columnwidth]{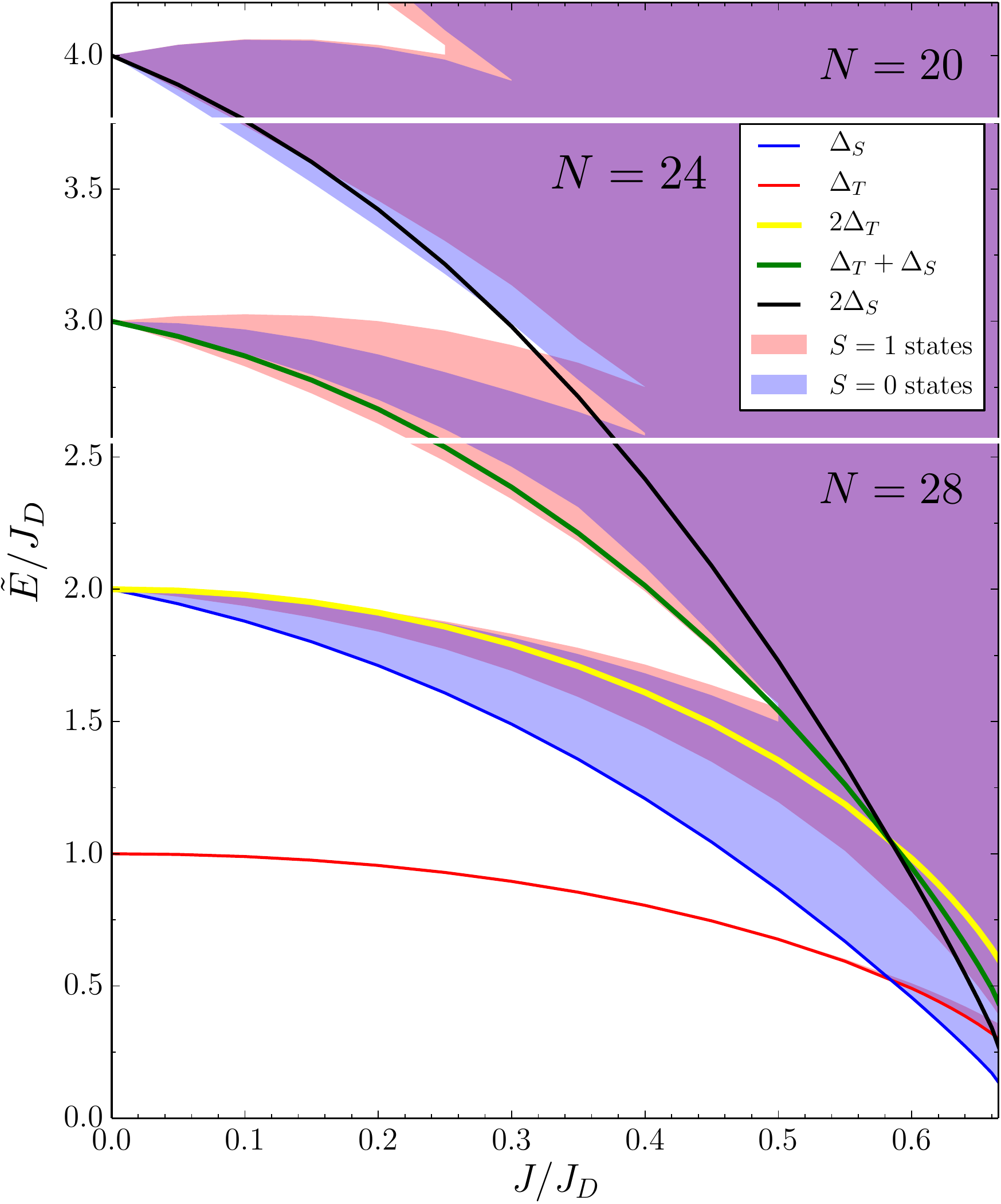}
  \caption{Low-lying energy spectrum of the Shastry-Sutherland model in 
    the $S = 0$ and $S = 1$ sectors, computed by ED using clusters of 
    sizes $N = 28$ (to ${\tilde E} = 2.6 J_D$), $N = 24$ (to $3.75 J_D$), and 
    $N = 20$ (to $4.2 J_D$) and shown over the full range of coupling ratios, 
    $J/J_D$, in the dimer-product phase. The excitation energy is denoted by 
    ${\tilde E} = E - E_{\rm GS}$, where $E_{\rm GS}$ is the energy of the ground 
    state. The shading represents regions with discrete but dense singlet 
    (blue) and triplet excitations (pink). The lowest pink line shows the 
    one-triplon sector, whose lower edge is $\Delta_T (J/J_D)$; the yellow 
    line shows the two-triplon scattering threshold, $2 \Delta_T (J/J_D)$. 
    The lower edge of the lower blue sector, which consists of singlet 
    bound states of two triplons, is $\Delta_S (J/J_D)$. We draw attention 
    to how few states in the three-triplon sector fall below the green line 
    ($\Delta_S + \Delta_T$) and in the four-triplon sector below the black 
    line ($2 \Delta_S$).}
  \label{fig:spec}
\end{figure}

Because the maximum of $\chi(T)$ is broad and flat, for an accurate 
characterization of the susceptibility we focus not on its position 
($T_{\rm max}^\chi$) but on the temperature, $T_{1/2}^\chi$ \cite{rus2}, at 
which $\chi(T)$ reaches half its maximum value during its rapid onset. 
Both quantities are shown in Fig.~\ref{fig:ph}(c), which indeed confirms 
significantly larger uncertainties in $T_{\rm max}^\chi$ than in $T_{1/2}^\chi$. 
$T_{1/2}^\chi$ tracks both $T_{\max}^C$ and $T_{1/2}^C$ rather closely throughout 
the range of Fig.~\ref{fig:ph}, albeit with a slightly slower decline very 
near the QPT. In this sense it functions as a comparable diagnostic of the 
falling energy scale in the system. Following the expectation that $\chi(T)$ 
depends on the gap to spinful excitations, in Fig.~\ref{fig:ph}(c) we compare 
this scale with the triplet gap, $\Delta_T (J/J_D)$, finding a constant of 
proportionality of approximately 0.22. 

Figure \ref{fig:ph} is a very specific diagnostic for the nature of the 
spectrum of low-lying energy levels, reflecting in particular a large number 
of states that become more nearly degenerate and are characterized by an 
ever-smaller (but always finite) gap as the system approaches the QPT. In 
Ref.~\cite{Wessel2018} we showed a schematic illustration of the energy 
spectrum of the Shastry-Sutherland model as a function of the coupling 
ratio $J/J_D$, based on Lanczos ED calculations to obtain the lowest 
levels of a cluster of size $N = 36$ sites. For more specific 
insight into the nature of the states revealed by the Lanczos procedure, 
here we have computed a much larger number of low-energy states, but for 
smaller clusters (of sizes up to $N = 28$). In Fig.~\ref{fig:spec} we gather 
these results to show the low-lying spectrum of the system in the $S = 0$ and 
$S = 1$ sectors over the full range of coupling ratios. We stress that, given 
the large number of states in these calculations, it is not easy to observe 
the discrete structure of the spectrum even in its lowest-lying regions, and 
as in Ref.~\cite{Wessel2018} we use shading to indicate the support of these 
excitations. 

In Fig.~\ref{fig:spec}, the thin red line at the bottom of the triplet 
spectrum is the elementary ``triplon'' excitation, which has energy ${\tilde 
E} = J_D$ at $J = 0$ and remains extremely narrow (non-dispersive) throughout 
the dimer-product phase due to the perfect frustration (Fig.~\ref{fig:sspd}). 
It is clear that significant numbers of two- and other multi-triplon states, 
in both the spin sectors shown, move to very low energies as the system 
approaches the QPT. Because the lowest states in the singlet sector must 
be of two-triplon origin, and these cross below the one-triplon energy around 
$J/J_D = 0.60$ (the exact value depends on $N$), it is equally clear that the 
physics of the Shastry-Sutherland model in the regime near the QPT is 
dominated by states one may classify as ``strongly bound.'' To specify the 
meaning of this term, the yellow line shows the quantity $2 \Delta_T (J/J_D)$, 
i.e.~twice the single-particle gap, which corresponds to the two-triplon 
scattering threshold, and near the QPT many states lie significantly below 
this energy. We comment that the separate ``families'' of excitations that can 
be traced back to independent two-triplon states at energy ${\tilde E} = 2 J_D$ 
at $J = 0$, and similarly for higher triplon numbers [Fig.~\ref{fig:spec}], 
exhibit a much larger dispersion across the Brillouin zone than do the 
one-triplon states, due to the smaller effects of geometric frustration 
on the propagation of bound states \cite{Knetter00,Totsuka2001,MiUeda03}. 
Although some authors have performed perturbative studies of the lowest-lying 
bound states of the Shastry-Sutherland model \cite{Knetter00}, we are not 
aware of previous attempts to count them systematically for comparison with 
thermodynamic properties, as we do next. 

\begin{figure}[t]
  \includegraphics[width=0.9\columnwidth]{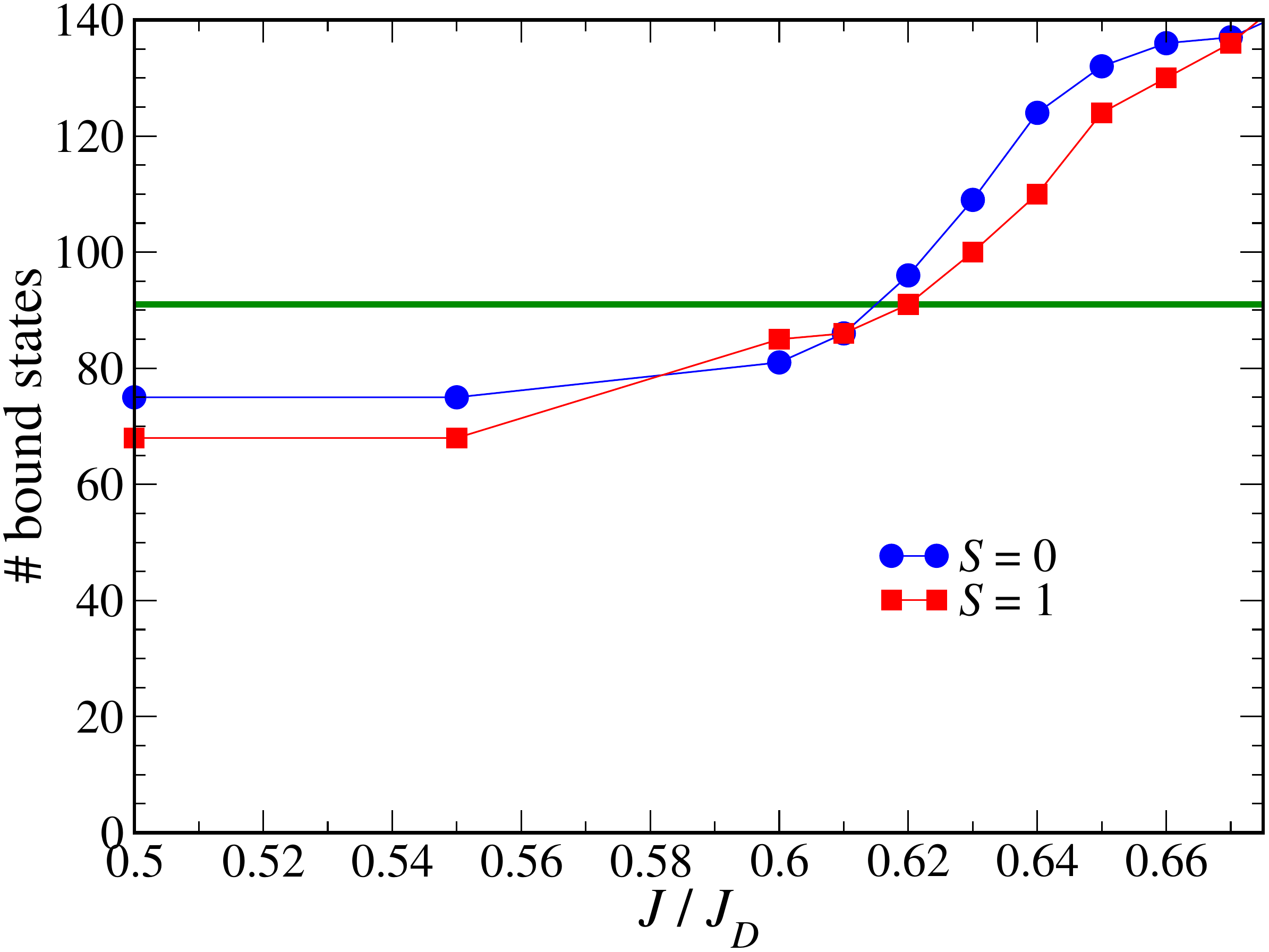}
  \caption{Numbers of low-lying singlet and triplet states on an $N = 28$ 
    lattice, shown as functions of $J/J_D$ in the dimer-product phase. Symbols 
    show ED data while connecting lines are guides to the eye. The horizontal 
    green line shows the number of independent two-triplon scattering states.}
  \label{fig:sc}
\end{figure}

Beyond illustrating significant binding-energy effects in the two- and 
higher-triplon sectors, Fig.~\ref{fig:spec} does not provide further 
information about the structure of the spectrum. As a step in this direction, 
Fig.~\ref{fig:sc} shows the numbers of singlet and triplet states lying below 
the threshold given by the yellow line in Fig.~\ref{fig:spec}. The number of 
triplet states excludes the $N/2$ states of the one-triplon band. For 
comparison, the expected number of two-triplon scattering states, 
${\textstyle \frac{1}{4}} N ({\textstyle \frac12} N - 1)$, is shown 
by the green line. Clearly the numbers of ``bound'' states in both sectors 
are significant fractions of the number of scattering states throughout the 
region $J/J_D \lesssim 0.55$. Further, the number of $S = 1$ states is 
comparable to that in the $S = 0$ sector, despite the fact that the energy 
gain is greater in the latter. As the QPT is approached, specifically once 
$J/J_D > 0.62$ for $N = 28$ [Fig.~\ref{fig:sc}], the numbers of low-lying 
states in both sectors increase well beyond the number of independent 
two-triplon states.

\begin{figure}[t]
  \includegraphics[width=0.9\columnwidth]{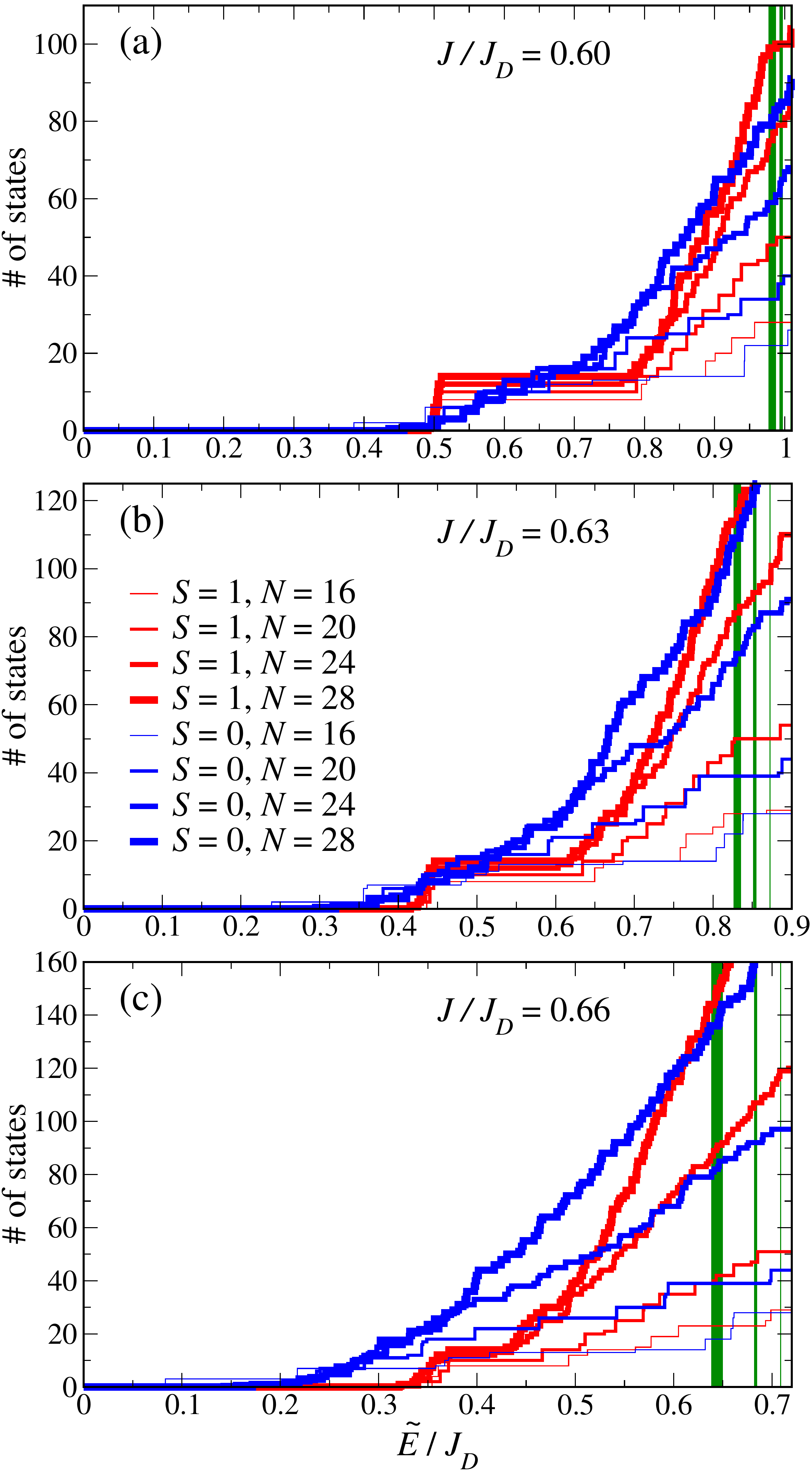}
  \caption{Total number of excited states lying below the energy $\tilde{E}$ 
    at coupling ratios (a) $J/J_D = 0.60$, (b) $J/J_D = 0.63$, and (c) $J/J_D
     = 0.66$. Blue lines indicate $S = 0$ excitations, red lines $S = 1$. 
    Increasing line widths indicate increasing system sizes, $N$, on which ED 
    calculations were performed. Vertical green lines denote the two-triplon 
    scattering thresholds for the corresponding system sizes.}
  \label{fig:cnt}
\end{figure}

Before discussing the nature of these states, for further information about the 
low-lying spectra we show in Fig.~\ref{fig:cnt} the total numbers of singlet 
(blue) and of triplet states (red) lying below a given energy, ${\tilde E}$, 
in ED calculations performed with four different system sizes for three fixed 
values of $J/J_D$ in the interval near the QPT. Here the $N/2$ one-triplon 
excitations are included in the triplet count, appearing as the plateau 
region in all the red lines directly above their onset. We note that all 
excitations with higher spin, $S \ge 2$, lie above the two-triplon scattering 
threshold for all $J/J_D$ values shown; while binding effects have also been 
observed in the $S = 2$ sector \cite{Corboz2015}, these are extremely small 
and are not detectable in clusters up to $N = 28$ due to finite-size effects. 

The evolution of this spectrum as a function of the coupling ratio contains 
two new messages beyond the ever-increasing number of very low-lying states 
shown in Fig.~\ref{fig:sc}. First, the steadily decreasing gap in the 
one-triplon sector can be expected to control the response to $\Delta S 
\ne 0$ processes over most of regime $0.60 \le J/J_D \le 0.66$, but very 
near the QPT it is overwhelmed by the high density of other triplet states 
that finally fall to this gap. Second, the dominant feature of the spectrum 
in this regime is the large number of low-lying singlet states, which are 
governed by an energy scale that in turn falls significantly more quickly 
with $J/J_D$ than the triplet gap. 

To discuss the nature of these states, and hence the consequences of our 
calculated spectra for the thermodynamic response functions computed in 
Sec.~\ref{tresults}, it is helpful to compare and contrast the 
Shastry-Sutherland model with the fully frustrated two-leg spin ladder. 
In this 1D model, some of us \cite{rus1,rus2} also found a striking  
thermodynamic response at low temperatures, which could be traced to a 
proliferation of low-energy multi-magnon bound states upon approaching a 
first-order QPT. While it is clear that the proliferation of low-lying states 
is the origin of the low-temperature properties of the Shastry-Sutherland 
model, there are some important differences between the 1D and 2D systems, 
most notably concerning the mechanism behind the formation of these states.

First, in the fully frustrated two-leg ladder, it was clear that these 
states were truly bound, meaning that they were intrinsic combinations of 
$n$ contributing triplons, with $n \ge 2$ becoming very large near the QPT. 
By contrast, in the Shastry-Sutherland system one may identify essentially 
all of the additional low-lying states, whose origin clearly lies in 
$n \ge 3$ triplons, with the expected scattering states of two singlet 
two-triplon bound states or of a singlet bound state and an elementary 
triplon. It is self-evident that when the energy of a singlet bound state 
of two triplons falls beyond the one-triplon gap, the lowest-lying scattering 
states of a pair of singlet bound states must fall below the two-triplon 
scattering threshold in the $S = 0$ sector, shown by the black line in 
Fig.~\ref{fig:spec}, and so does the scattering state of a singlet bound 
state and an elementary triplon in the $S = 1$ sector, shown by the green 
line in the same figure. Thus the proliferation of low-lying excitations in 
the regime $J/J_D > 0.6$ can, due to the differences in state-counting in 2D, 
be driven (almost) completely by the strength of the two-triplon binding 
effect in the singlet sector. It is for this reason that we do not refer to 
all of the low-lying states we observe here as ``bound states.'' Concerning 
the possibility of true multi-triplon binding effects, as noted above we may 
comment only that, if present, these are sufficiently weak that we are unable 
to detect them.

Second, at the QPT in the fully frustrated ladder, the bound states cluster 
strongly at a single energy value of approximately 85\%~of the one-triplon 
gap \cite{rus1,rus2}, but in the Shastry-Sutherland model we observe no such 
sharp structures in the density of states. Specifically, although we find for 
$J/J_D = 0.66$ that low-lying singlet states far outnumber the triplets at 
any energy $\tilde{E} < 0.5 J_D$ [Fig.~\ref{fig:cnt}(c)], there is no sign 
of a discrete set of degenerate levels, at any energy, that could be related 
directly to the specific-heat peak. Instead the number of excited states
continues to rise with ${\tilde E}$, in fact at an increasing rate, such that 
some very high densities of states may be found at energies of order $0.5 J_D$ 
above the lowest ones in each panel of Fig.~\ref{fig:cnt}. The distinguishing 
feature of these broadly distributed states, at least in the singlet sector, 
is that their binding effects are energetically stronger than in the 1D 
system. Although finite-size effects prevent us from reaching the QPT 
($J/J_D = 0.675$), at $J/J_D = 0.66$ the singlet two-triplon bound states 
have already descended to only 54\%~of the one-triplon gap (on the 
$N = 28$ cluster). Thus the low-temperature peak in $C(T)$ observed in 
Sec.~\ref{tresults} emerges at a relatively lower energy scale than in 
the fully frustrated ladder. Certainly the possibility of ascribing our 
results to only one energy scale is consistent with our discovery in 
Fig.~\ref{fig:ph}(a) that the peak shapes are identical. In this regard, 
the strikingly narrow specific-heat peaks near the QPT are not in fact 
anomalous, in that their relative width is constant and their appearance 
is a consequence of this one decreasing energy scale. Regarding the 
susceptibility, the decrease of the one-triplon energy scale in 
Fig.~\ref{fig:cnt} appears to provide a reasonable account of the decrease 
in the ``onset'' temperature indicator $T_{1/2}^\chi$ [Fig.~\ref{fig:ph}(c)] 
over the entire range of $J/J_D$, while the steep fall in energy of the other 
low-lying triplet states may be responsible for the increasingly abrupt 
(``square'') turnover at the maximum in $\chi(T)$. 

In summary, our detailed ED investigation of the low-energy spectrum 
[Figs.~\ref{fig:spec}, \ref{fig:sc}, and \ref{fig:cnt}] reveals the rapid 
descent in energy of a multitude of singlet and triplet states as the QPT 
is approached. Despite the narrow nature of the $C(T)$ peak, these states 
do not converge to a single characteristic energy in either sector. In 
contrast to the $n$-triplon bound states of the fully frustrated ladder near 
its QPT, the energetics of the low-lying states of the Shastry-Sutherland 
model are dominated by the binding energy of two triplons into a net singlet. 
Because this local bound state involves two neighboring orthogonal dimers 
\cite{Knetter00,Totsuka2001}, it has little or no relation with the 
plaquette singlets that control the physics on the other side of the QPT 
(Fig.~\ref{fig:sspd}), and hence one may conclude that the first-order 
transition involves a complete rearrangement of spin correlations in the 
system. 

\subsection{Numerical Methods}

We comment only briefly on the limitations and prospects for improvement 
of our numerical techniques. In a sense the Shastry-Sutherland model near 
the QPT presents an ideal test-bed for any method, in that the low-$T$ peak 
in the specific heat becomes progressively narrower, and thus more challenging 
to reproduce, as the QPT is approached. In our ED and TPQ calculations, 
it is both clear and completely unsurprising that larger cluster sizes improve 
the ability of both methods to capture this peak (and the corresponding 
susceptibility shoulder). Whereas the thermodynamic response at coupling 
ratio $J/J_D = 0.60$ could not be reproduced with complete accuracy by 
$N = 20$ ED, or indeed $N = 28$ Lanczos ED (Fig.~\ref{fig:sspd06}), we may 
assert that $J/J_D = 0.63$ can be treated accurately by TPQ with $N \le 40$ 
(Fig.~\ref{fig:sspd063}). 

However, the limits to TPQ appear to be reached at $J/J_D > 0.65$, where 
our calculations are no longer able to capture the anomalously narrow $C(T)$ 
peak that forms in this regime (Fig.~\ref{fig:sspd066}). With reference to 
our discussion of Sec.~\ref{bla}A, one may only speculate that the physics 
of this peak resides in scattering states of the singlet two-triplon bound 
states whose fully developed real-space extent begins to exceed this cluster 
size very close to the QPT. As noted in Sec.~\ref{ttpq}, reaching larger 
cluster sizes with acceptable statistical errors is not possible with present 
computing power. Similarly TPQ, like ED, is not limited by algorithmic 
complexity, and indeed our present calculations already exploit all available 
spin and spatial (cluster) symmetries, implying that more complex Hamiltonians 
would be subject to stricter limits on $N$. 

By contrast, our iPEPS calculations have relatively modest CPU and memory 
requirements. To date they have been run only on a single node, and thus 
larger-scale computations would certainly become feasible through the future 
development of a highly parallelized code. In comparison with more mature 
numerical methods, tensor-network calculations remain in their relative 
infancy, and as a result retain significant scope for algorithmic improvement. 
In current implementations, both CPU and memory requirements increase as high 
powers of the bond dimension, $D$; although most developments focus on reducing 
these exponents, it is not proven beyond doubt that large $D$ is the only route 
to an accurate tensor-network description. As one example, we comment that 
higher accuracies can be achieved for the same $D$ value within our present 
calculations by the use of disentangling gates acting on the ancillas 
\cite{hauschild18,czarnik19}. 

Nevertheless, as demonstrated in Sec.~\ref{tipeps}, the influence of the 
other variables in the iPEPS method (Trotter step, CTM boundary bond dimension, 
type of update) is small in comparison with the effects of $D$. Physically, 
there is at present no known means of relating $D$ as a figure of merit in 
finite-$T$ (i.e.~thermal-state purification) iPEPS implementations with $D$ 
in a ground-state iPEPS calculation. As with the problem of which absolute 
value of $D$ is able to capture the physics of any given Hamiltonian 
(Sec.~\ref{tipeps}), the most reliable approach to date is the empirical 
one of increasing $D$ and monitoring changes and convergence. As we have 
shown in Secs.~\ref{tipeps} and \ref{tresults} (Fig.~\ref{fig:ndc}), for the 
present problem we have indeed reached a well-defined limiting regime of $D$, 
even at $J/J_D = 0.66$. Thus one may conclude that the Shastry-Sutherland 
model does not present a physical system where calculations are limited by 
the ability to capture long-ranged entanglement, as may perhaps be expected 
when the ground state is a product state, all the excitations are gapped, 
and the nearby QPT is first-order.

\subsection{Experiment}

In the light of the data and insight obtained from applying our two advanced 
numerical methods to the thermodynamics of the Shastry-Sutherland model, we 
revisit the experimental situation in SrCu$_2$(BO$_3$)$_2$. We note that 
low-lying singlet excitations in SrCu$_2$(BO$_3$)$_2$ were documented by 
Raman scattering soon after the discovery of the material \cite{Lemmens00}, 
and later that the same method was used to study the low-lying spectrum 
in an applied magnetic field \cite{Gozar05}. The triplet excitations were 
studied at first by electron spin resonance (ESR) \cite{Nojiri99} and 
inelastic neutron scattering \cite{Kageyama2000b}. Subsequent high-field 
ESR measurements were used to demonstrate the presence of two-triplon bound 
states \cite{Nojiri03}, while a recent application of modern neutron 
spectroscopy instrumentation studied the lowest bound triplet mode, and 
its response in a magnetic field, in unprecedented detail \cite{McClarty17}. 
The evolution of the triplet spectrum has been studied under pressure 
\cite{Zayed2017}, which as noted in Sec.~\ref{sintro} acts to increase the 
ratio $J/J_D$ through the first-order QPT out of the dimer-product phase. 
Although an analogous study of the thermodynamic properties has appeared in 
parallel with the completion of the present study \cite{Guo2019}, only one 
dataset for $C(T)/T$ is shown at a pressure between ambient and the QPT; the 
low-$T$ peak moves down by approximately 25\% at this pressure and undergoes 
a definite increase in sharpness. Around the QPT, extremely sharp $C(T)/T$ 
peaks are found in some, but not all, samples. We observe that the numerical 
modeling that accompanies these measurements does not advance the state of 
the art beyond that of Ref.~\cite{MiUeda00}.

\begin{figure}[t]
  \includegraphics[width=\columnwidth]{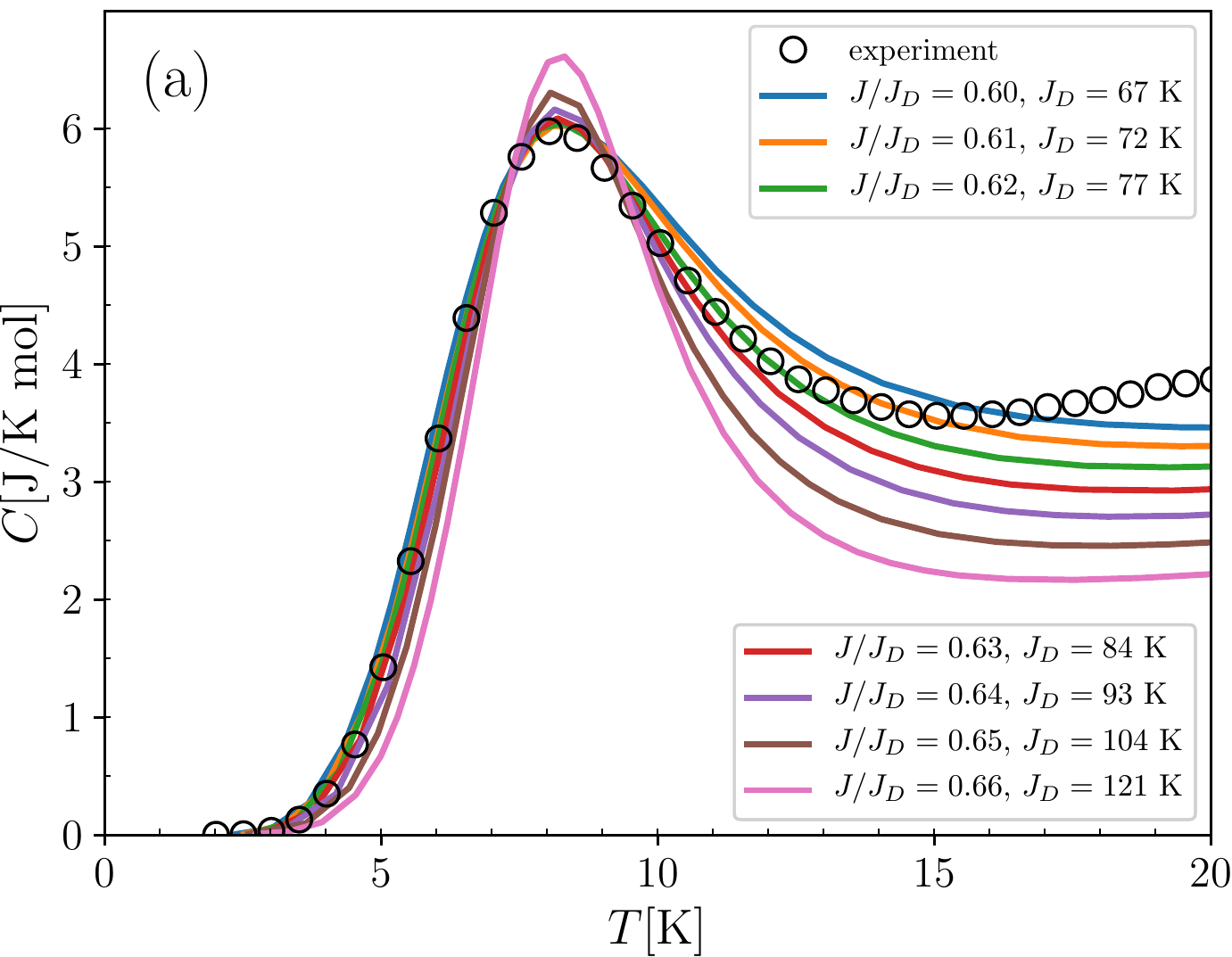}
  \includegraphics[width=\columnwidth]{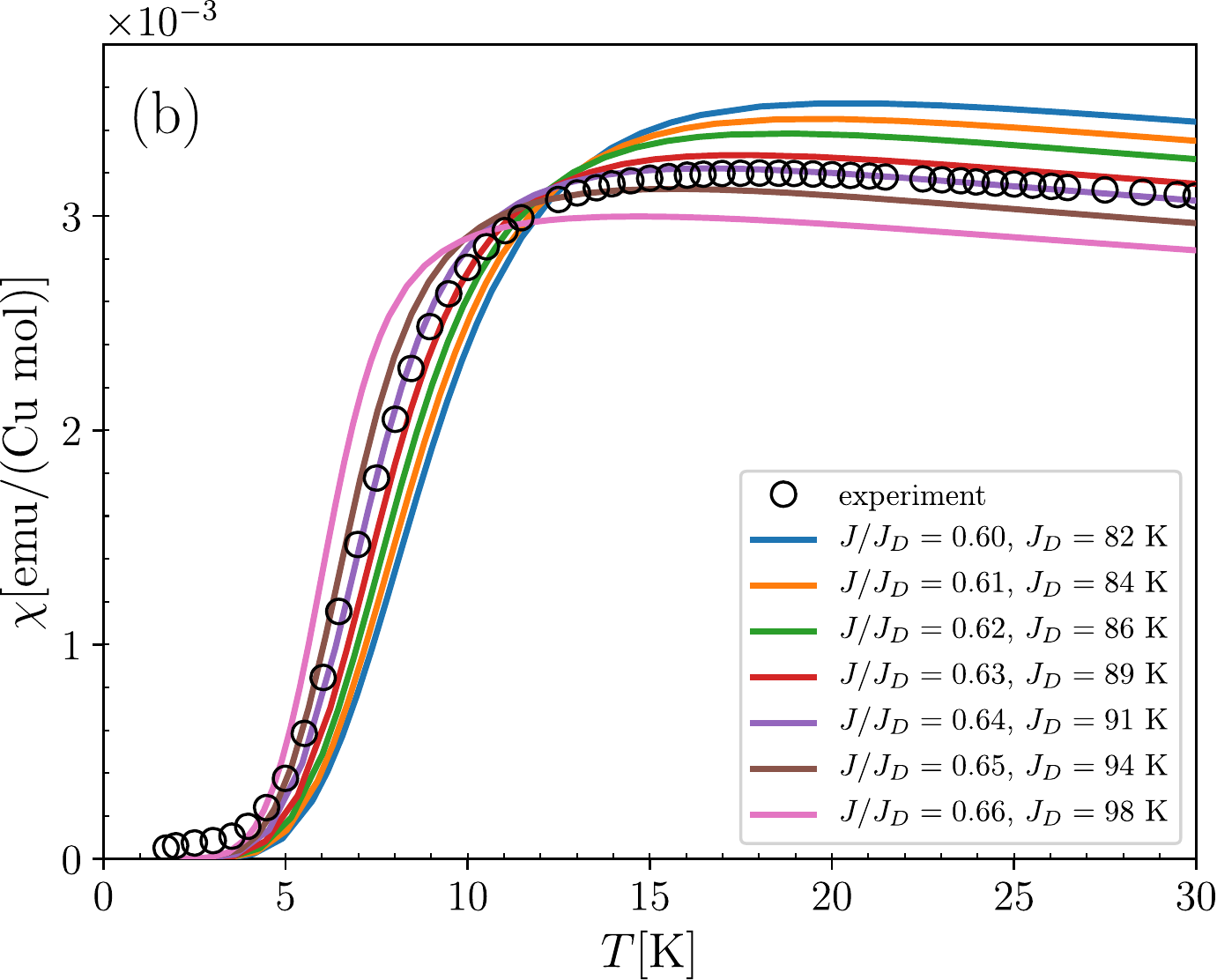}
  \caption{$D = 18$ iPEPS calculations of the magnetic specific heat (a) and 
    susceptibility (b) performed for coupling ratios 0.60, 0.61, \dots, 0.66 
    and for $J_D$ values dictated by experiment. Specific-heat data in panel 
    (a) were taken from Ref.~\cite{Kageyama2000} and a phonon contribution 
    of $0.5\,T^3\,{\rm mJ}/({\rm mol}\,{\rm K}^4)$ was subtracted. 
    Susceptibility data in panel (b) were taken from Ref.~\cite{Kageyama1999b}
    and a $g$-factor of 2.28 was used.}
  \label{expipeps}
\end{figure}

We comment first that a rather accurate understanding of the high-temperature, 
and indeed high-field, properties of SrCu$_2$(BO$_3$)$_2$ can be obtained using 
the Shastry-Sutherland model with a coupling ratio close to $J/J_D = 0.63$, 
and it is the remaining mystery surrounding the low-$T$ response that we 
address here. We use our $D = 18$ iPEPS results as a consistent indicator of 
the evolution of the specific heat and susceptibility with the coupling ratio, 
and in Fig.~\ref{expipeps} we show both quantities for all $J/J_D = 0.60$, 
0.61, \dots, 0.66. For the specific heat [Fig.~\ref{expipeps}(a)] we compare 
our results with the data of Ref.~\cite{Kageyama2000} for SrCu$_2$(BO$_3$)$_2$, 
from which we have subtracted a phonon contribution $C_{\rm ph} = \gamma \,T^3$ 
with $\gamma = 0.5$ mJ/(mol K$^4$). For the susceptibility 
[Fig.~\ref{expipeps}(b)] we compare our results with the data of 
Ref.~\cite{Kageyama1999b}, using a $g$-factor of $g_c = 2.28$ \cite{Nojiri99}. 
The optimal values of $J_D$ required to reproduce the position and shape of 
the low-$T$ peak in $C(T)$ and to reproduce the rapid upturn and maximum 
height of $\chi(T)$ are determined separately; in both cases they turn out 
to be strong functions of the coupling ratio, and hence can be used for 
accurate estimation of this energy scale. 

In Fig.~\ref{expipeps}(a) it is safe to say that the best account of the 
low-$T$ peak in $C(T)$ is given by $J/J_D = 0.62$, with $J_D = 77$ K. Somewhat 
surprisingly, it is not necessary to exploit to its limits our abilities to 
resolve a very narrow peak in $C(T)$, as would be the case at higher coupling 
ratios. Because $J_D$ is constrained by many other aspects of the experimental 
data for SrCu$_2$(BO$_3$)$_2$, its strong dependence on $J/J_D$ serves as 
a strict constraint on the coupling ratio. Turning to Fig.~\ref{expipeps}(b), 
the best account of $\chi(T)$ is given by $J/J_D = 0.64$, with $J_D = 91$ K. 
Here it is safe to say that the fits in both panels of Fig.~\ref{expipeps} 
are difficult to reconcile, certainly at the experimentally determined value 
of $g_c$, and as a result we can conclude that we have found the limits of the 
Shastry-Sutherland model when applied to the SrCu$_2$(BO$_3$)$_2$ problem. 
It is known that the appropriate spin Hamiltonian for this material contains 
additional terms, most notably an interlayer coupling, whose magnitude has 
been estimated at 9\% of $J_D$ \cite{MiUeda03}, and Dzyaloshinskii-Moriya 
interactions both within and normal to the plane of the system, whose 
magnitudes have been estimated \cite{Nojiri03} for both components at 3\% 
of $J_D$. Because it is not the purpose of our present analysis to address 
these details, for further comparison in the present context we restrict 
our attention to establishing the optimal Shastry-Sutherland fit for 
SrCu$_2$(BO$_3$)$_2$, to which end we adopt the compromise coupling ratio 
$J/J_D = 0.63$ and find that the corresponding optimal $J_D$ is 87 K.

\begin{figure}[t]
  \includegraphics[width=\columnwidth]{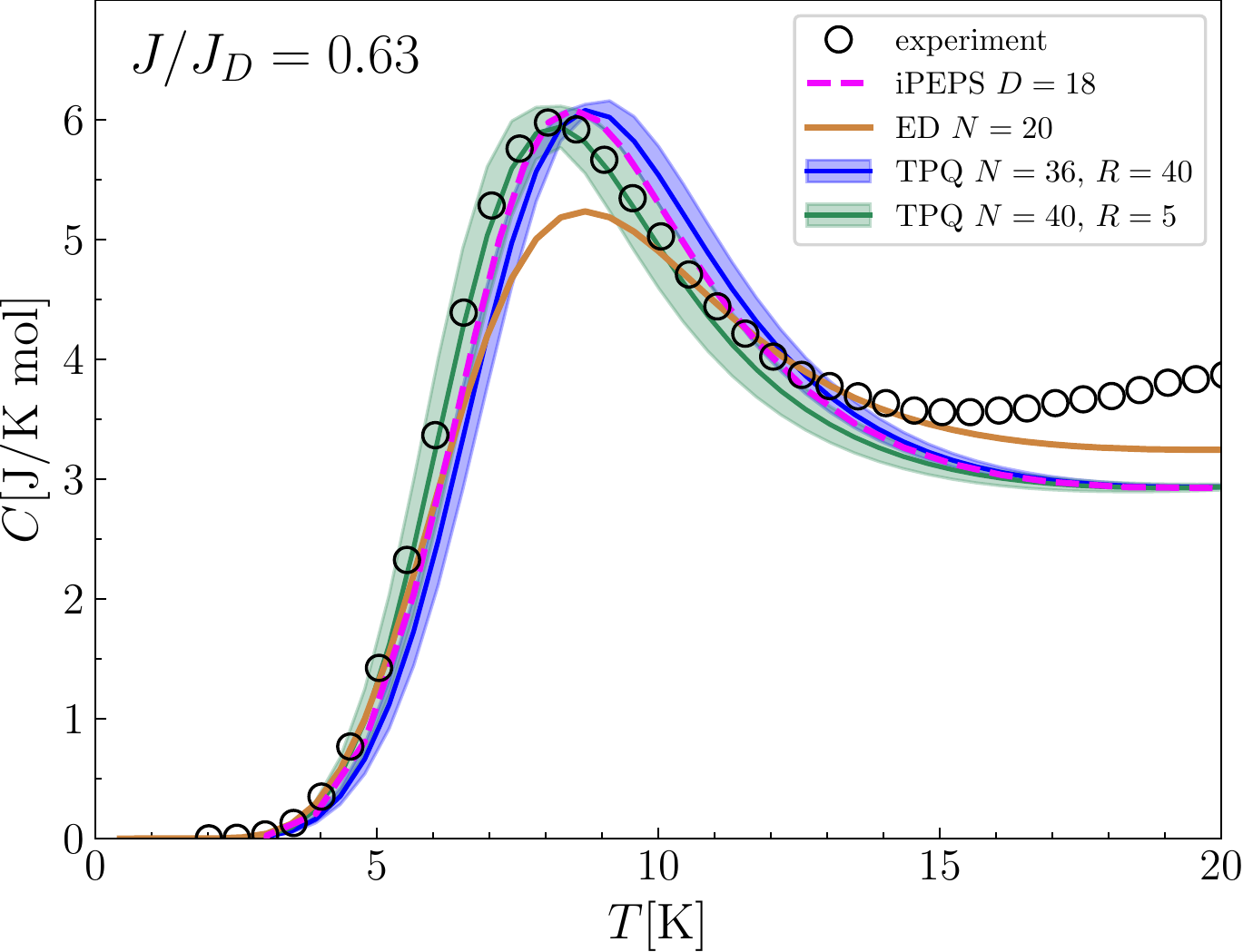}
  \caption{Comparison of experimental measurements of the magnetic specific 
    heat, taken from Ref.~\cite{Kageyama2000} and with a phonon 
    contribution $0.5\,T^3\,{\rm mJ}/({\rm mol}\,{\rm K}^4)$ subtracted, 
    to numerical calculations performed by ED, TPQ, and $D=18$ iPEPS for 
    $J/J_D = 0.63$ with $J_D = 87$~K.}
  \label{fig:expcc}
\end{figure}

In Fig.~\ref{fig:expcc} we show a quantitative comparison between the results 
of Sec.~\ref{tresults}, specifically Fig.~\ref{fig:sspd063}, and the $C(T)$ 
data of Ref.~\cite{Kageyama2000}. We observe that our best TPQ results ($N = 
40$) and our iPEPS results separately provide quantitatively excellent 
fits to the measured data. Our conclusion that such a fit is optimized with 
the coupling ratio $J/J_D = 0.63$ is in accordance with the suggestion, made 
on the basis of small-system ED calculations, of Ref.~\cite{MiUeda00}. Further, 
our determination of the coupling constant, $J_D = 87$~K, from the precisely  
known position, $T^C_{\max}$ [Fig.~\ref{expipeps}(a)], of the $C(T)$ peak, is 
also close to the parameters deduced 20 years ago. As noted above, while 
these constants may also be constrained from high-$T$ and magnetization 
information, our results demonstrate finally that they are consistent with 
the hitherto mysterious low-$T$ behavior. 

\begin{figure}[t]
  \includegraphics[width=\columnwidth]{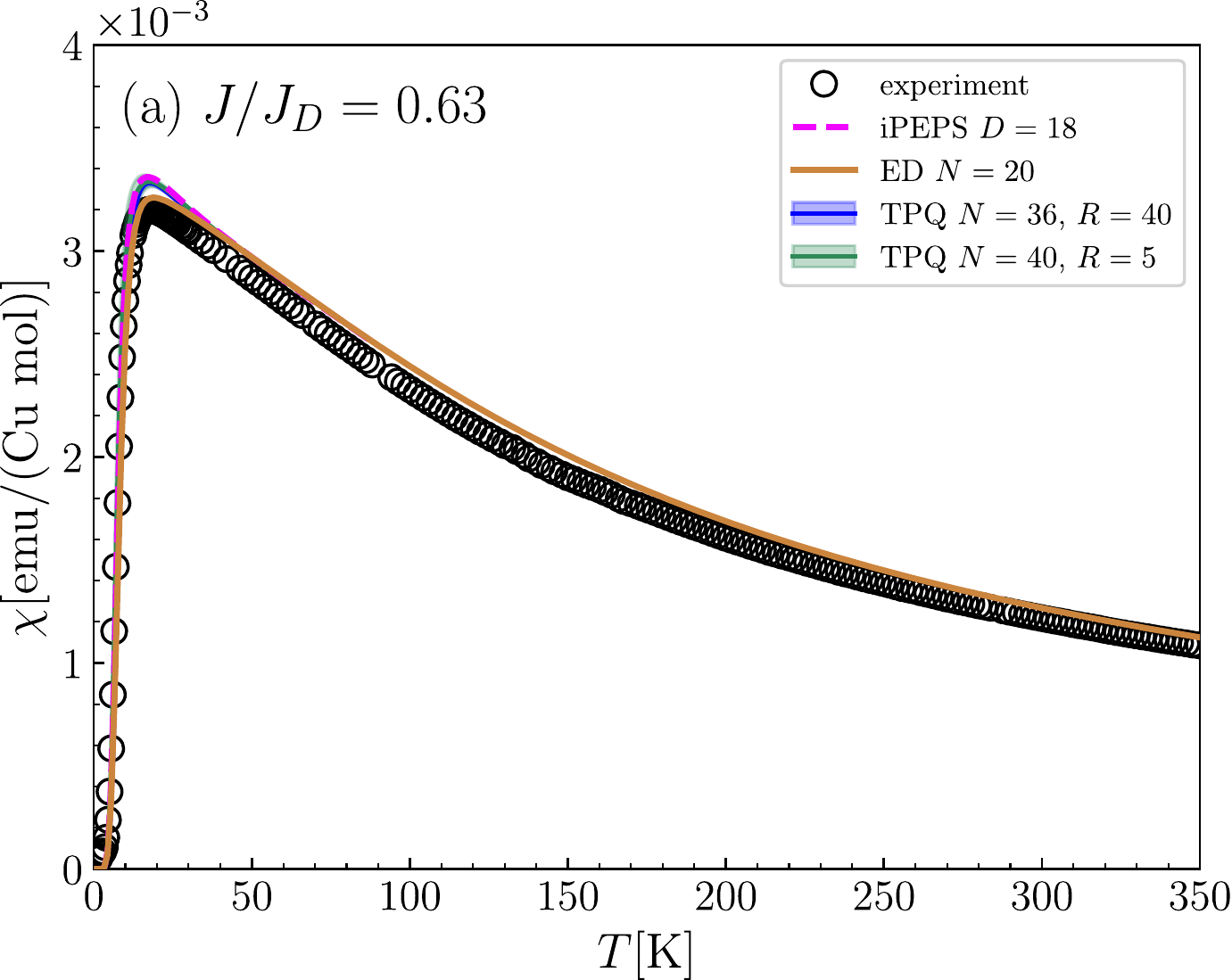}
  \includegraphics[width=\columnwidth]{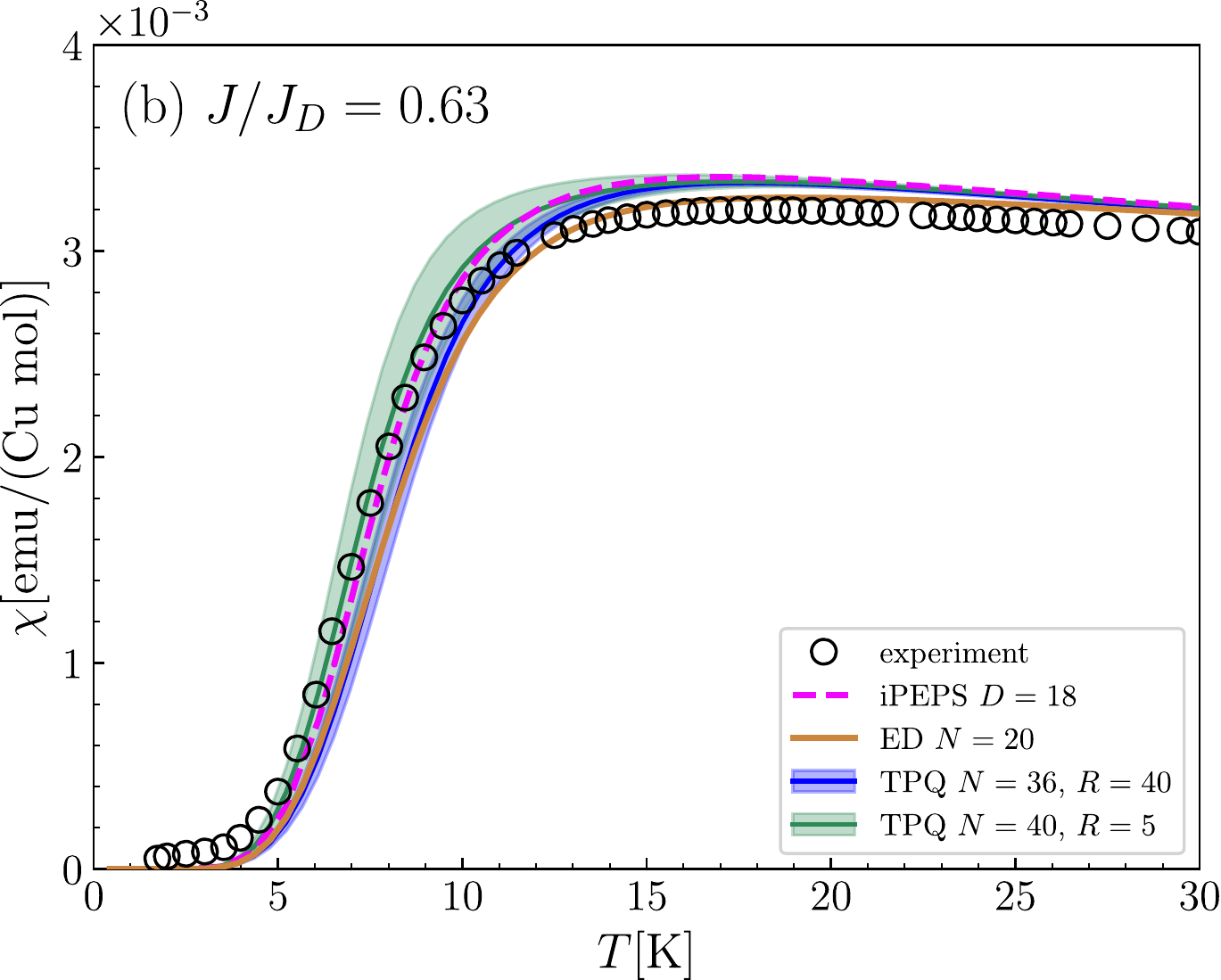}
  \caption{(a) Comparison of experimental measurements of the magnetic 
    susceptibility, taken from Ref.~\cite{Kageyama1999b}, to numerical 
    calculations performed by ED, TPQ, and $D=18$ iPEPS for $J/J_D = 0.63$. 
    We use $g = 2.28$ and $J_D = 87$~K. (b) Detail of the comparison at 
    low temperatures.}
  \label{fig:expcchi}
\end{figure}

We stress that, despite reaching the previously unattainable goal of 
reproducing the true shape of the low-$T$ peak in $C(T)$, indeed for coupling 
ratios even higher than $J/J_D = 0.63$, we do not find that a sharper peak is 
required to explain experiment (our results at $J/J_D \ge 0.64$ tend to exceed 
the experimental peak height). Thus we deduce that, to the extent that the 
Shastry-Sutherland model offers an acceptable account of the physics of 
SrCu$_2$(BO$_3$)$_2$ (i.e.~to the extent that 3D and Dzyaloshinskii-Moriya 
interactions may be neglected), the coupling ratio is $J/J_D = 0.63 \pm 
0.01$. We have based this estimate on reproducing the shape of the low-$T$ 
peak in $C(T)$; if we consider the dip region at higher temperatures, 
Fig.~\ref{fig:expcc} shows a significant discrepancy between experiment 
and our best peak fits, but because the question of subtracting the phonon 
contribution becomes more important here it is difficult to ascribe the same 
quantitative reliability to our fits. 

In Fig.~\ref{fig:expcchi}, we perform the same exercise by showing a 
quantitative comparison between our results (Sec.~\ref{tresults}) and 
the data of Ref.~\cite{Kageyama1999b} for the magnetic susceptibility of 
SrCu$_2$(BO$_3$)$_2$. Again we use the value $J_D = 87$ K along with the 
$g$-factor $g_c = 2.28$ noted above. We show the entire temperature 
range in Fig.~\ref{fig:expcchi}(a), which emphasizes the very abrupt 
peak and the reliability of our model at all high temperatures.
In Fig.~\ref{fig:expcchi}(b) we focus on the low-temperature regime, 
where again the half-height, $T_{1/2}^\chi$, admits only a rather narrow 
range of coupling ratios. Thus our improved computational methods confirm 
both the extent to which SrCu$_2$(BO$_3$)$_2$ can be approximated by a pure 
Shastry-Sutherland model and the efficacy of the parameter estimates made 
two decades ago on the basis of significantly inferior numerical technology. 

\section{Summary and Perspectives}
\label{wrap}

We have introduced two advanced numerical techniques, the methods of typical 
pure quantum (TPQ) states and infinite projected entangled-pair states (iPEPS), 
for computing the thermodynamic properties of quantum magnets. Both approaches 
are unaffected by frustration and we apply them to compute the magnetic 
specific heat and susceptibility of the Shastry-Sutherland model in its 
dimer-product phase near the first-order quantum phase transition (QPT) to 
a plaquette phase. This challenging region of coupling ratios has remained 
entirely impervious to meaningful analysis by any previous numerical 
techniques, and now we are able to reveal why. The specific heat develops 
an increasingly narrow low-temperature peak, which is separated from the 
broad hump due to local processes by an emerging dip. This peak is caused 
by a proliferation of low-lying energy levels whose origin lies in scattering 
states of multiple two-triplon bound states, and its anomalously low energy 
scale moves ever lower (albeit not to zero) as the QPT is approached. The 
analogous feature in the magnetic susceptibility is an increasingly rapid 
rise at an ever-lower temperature, followed by an abrupt maximum. This 
physics is a consequence of both high frustration and the proximity to a 
QPT, and its manifestations have been discussed previously in highly 
frustrated 1D systems, but now we have captured its realization in 2D.

Our methods and benchmarks offer very wide scope for immediate application. 
The list of open problems in frustrated quantum magnetism is long, and we 
await with interest definitive results, meaning for extended or spatially 
infinite systems, for the thermodynamic response of the triangular, kagome, 
and $J_1$-$J_2$ square lattices. The ability to fingerprint the energy 
spectrum at the level of thermodynamic properties will provide a new 
dimension of physical understanding, and in some cases may even yield new 
insight into the ground state. Away from the classic problems of geometrical 
frustration in Heisenberg models, there is a pressing need for thermodynamic 
information to characterize many other types of frustrated system, including 
spin-ice materials, candidate Kitaev materials, coupled anisotropic (Ising 
and XY) chain and planar materials, candidate chiral spin liquids, materials 
with Dzyaloshinskii-Moriya interactions, topological magnets, and certain 
skyrmion systems. 

Technically, neither of our methods presents any barrier to the application 
of a magnetic field, which could be used to control the relative singlet and 
triplet gaps and create qualitative changes in the thermodynamic response in 
many of the above types of system. For the SrCu$_2$(BO$_3$)$_2$ problem, the 
next step enabled by the capabilities developed here is clearly to follow 
the evolution of the system to and through the pressure-induced QPT into 
the plaquette phase \cite{Zayed2017,Guo2019}. This process presents a clear 
need for the investigation of thermodynamic properties, even if the physics 
of the material beyond the dimer-product phase is critically dependent on 
additional Hamiltonian terms beyond the Shastry-Sutherland model. 

\begin{acknowledgments}

We thank S. Capponi, P. Lemmens, J. Richter, Ch. R\"uegg, J. Schnack, R. 
Steinigeweg, A. Tsirlin, and B. Wehinger for helpful discussions. This work 
was supported by the Deutsche Forschungsgemeinschaft (DFG) under Grants 
FOR1807 and RTG 1995, by the Swiss National Science Foundation (SNF), and by 
the European Research Council (ERC) under the EU Horizon 2020 research and 
innovation programme (Grant No.~677061). The Flatiron Institute is a division 
of the Simons Foundation. SW thanks the IT Center at RWTH Aachen University and 
the JSC J\"ulich for access to computing time through JARA-HPC. AH is grateful 
for the use of the osaka cluster at the Universit\'e de Cergy-Pontoise. 

\end{acknowledgments}

\bibliography{ssd.bib}

\end{document}